\documentclass[jcp,onecolumn,preprint,floatfix]{revtex4-1}

\usepackage{graphics} 
\usepackage{graphicx} 
\usepackage{color} 
\usepackage{bm} 
\usepackage{bbm}
\usepackage[dvipsnames]{xcolor,colortbl}
\usepackage{amsfonts}
\usepackage[normalem]{ulem}  

\definecolor{DarkYellow}{RGB}{80, 80, 0}

\usepackage{amsmath,amssymb,graphicx}
\usepackage{amsbsy}
\usepackage{latexsym}
\usepackage{xcolor}
\usepackage{graphicx}
\usepackage{psfrag}
\usepackage[normalem]{ulem}
\usepackage{bm}
\usepackage{hyperref}
\usepackage[SF,sf]{subfigure}
\usepackage[nottoc]{tocbibind}
\hypersetup{colorlinks=true, linkcolor=red, citecolor=blue, urlcolor=blue}
\usepackage{float}
\usepackage{multirow}
\usepackage{caption}
\begin{document} 

\title{Fingerprints of ordered self-assembled structures in the liquid phase of a hard-core, square-shoulder system}

\author{Michael Wassermair$^{1,2}$}
\author{Gerhard Kahl$^{1}$}
\author{Roland Roth$^{3}$} 
\author{Andrew J.~Archer$^{2}$}

\affiliation{$^1$Institut f\"ur Theoretische Physik, TU Wien, Wiedner Hauptstra{\ss}e 8-10, A-1040 Vienna, Austria}
\affiliation{$^2$Department of Mathematical Sciences and Interdisciplinary Centre for Mathematical Modelling, Loughborough University, Loughborough LE11 3TU, United Kingdom}
\affiliation{$^3$Institute for Theoretical Physics, University of T\"ubingen, D-72076 T\"ubingen, Germany}


\begin{abstract}
We investigate the phase ordering (pattern formation) of systems of two-dimensional core-shell particles using Monte-Carlo (MC) computer simulations and classical density functional theory (DFT). The particles interact via a pair potential having a hard core and a repulsive square shoulder. Our simulations show that on cooling, the liquid state structure becomes increasingly characterised by long wavelength density modulations, and on further cooling forms a variety of other phases, including clustered, striped and other patterned phases. In DFT, the hard core part of the potential is treated using either fundamental measure theory or a simple local density approximation, whereas the soft shoulder is treated using the random phase approximation. The different DFTs are bench-marked using large-scale grand-canonical-MC and Gibbs-ensemble-MC simulations, demonstrating their predictive capabilities and shortcomings. We find that having the liquid state static structure factor $S(k)$ for wavenumber $k$ is sufficient to identify the Fourier modes governing both the liquid and solid phases. This allows to identify from easier-to-obtain liquid state data the wavenumbers relevant to the periodic phases and to predict roughly where in the phase diagram these patterned phases arise.
\end{abstract}

\date{\today}

\maketitle

\section{Introduction}
Systems of particles interacting pairwise via purely repulsive interaction potentials can exhibit a surprisingly rich phase behaviour. The particles that we consider here have hard impenetrable cores, surrounded by a penetrable corona. The resulting pair potential between the particles exhibits a hard core of diameter $\sigma$, beyond which is a repulsive shoulder of range $\lambda\sigma$, with energy penalty $\epsilon>0$ for overlapping coronas. The range parameter $\lambda>1$ depends on the extent of the coronas. Here, we treat the system as being two-dimensional (2D); the three-dimensional analogue is also very interesting, but beyond the scope of the present study. An experimental {realisation} of the 2D model is the system studied in Ref.~\cite{vogel2012ordered}, consisting of metallic nanoparticles with a corona of attached polymers. These particles are then adsorbed at a water-air interface, rendering the system effectively 2D. Other such systems include microgels \cite{geisel2014compressibility, rey2016isostructural, Volk2019} or various types of colloidal spheres decorated with polymer-chains \cite{rey2017anisotropic, Menath2021, Ickler2022}. These core-shoulder systems are of fundamental importance, because understanding and predicting the phase behaviour provides a crucial benchmark challenge for computer simulations and theories for interacting classical many-body systems. They are also of practical importance because the colloidal structures that self-assemble promise useful optical and/or material properties \cite{rey2017anisotropic}.

Core-shoulder systems do not exhibit gas-liquid phase separation, due to the lack of attractive interactions. Nonetheless, complex phase ordering can occur, depending on the range of the shoulder potential. Typically, they form cluster, stripe and hole phases at intermediate densities, as long as the shoulder repulsion is strong enough (or, equivalently, when the temperature is low enough) \cite{malescio2003stripe}, which at first sight can appear like the generic pattern formation exhibited by systems with competing interactions \cite{seul1995domain}. However, at higher densities and lower temperatures, a plethora of different ordered phases can arise, depending on the density, temperature, and range of the shoulder repulsion $\lambda$, often with a sensitive dependence on the precise model parameter values. Belying the simplicity and isotropic nature of the interactions, the structures that can form range from the clusters and stripes already mentioned, to various anisotropic structures \cite{Glaser2007, Dobnikar_2008, Fornleitner_2008, Fornleitner_2010, pattabhiraman2017formation, Menath2021} and even quasicrystals of various different symmetries \cite{Dotera2014, ziherl2016geometric, Dotera2017}.

The notion that softening the repulsive part of the interparticle pair potential can lead to the appearance of additional phase transitions goes back to Hemmer and Stell \cite{hemmer1970fluids}, who investigated systems of {hard-core} particles ``softened'' with a weakly attractive potential. Later Jagla introduced a more general class of two length-scale models \cite{jagla1998phase, Jagla1999} with a rich phase behaviour, {popularising} the topic. Such potentials have even been used to model liquid-liquid transitions in water \cite{Strekalova2012}. Here, we report results for systems having a shoulder potential that is step-like. However, various soft-shoulder analogues have been considered, such as exponential and tanh-shaped shoulders, or ramp-like potentials \cite{jagla1998phase, Jagla1999, Somerville2020}. Our general approach is relevant to this whole class of models.

Our focus here is to identify features of the liquid state that are indicators or precursors of the complex phase behaviour that arises at lower temperatures.
In other words, to allow for easy navigation of the phase diagram, we identify the features to look for in the liquid state that are associated with the anisotropic phases.
To describe the structural properties of our system we have on the one hand performed extensive Monte Carlo (MC) based simulations, including both grand-canonical ensemble MC (GCMC) and Gibbs ensemble MC (GEMC) \cite{Allen2017,frenkel_smit} and on the other hand we have used various complementary theoretical approaches, all based on classical density functional theory (DFT) \cite{Evans1979, hansen13, Evans_2016}.

A key part of our approach is to use DFT to determine the major density modes (i.e.\ the characteristic wavelengths) of the liquid state structure.
We use two different DFT approximations.
The first is a simple DFT that has the advantage of providing simple analytic expressions for many of the quantities of interest, while the second is a more sophisticated (fundamental measure theory \cite{hansen13} based) DFT, that still provides semi-analytic results for some quantities.
We compare results from the two DFTs and benchmark all predictions against the results from the MC simulations.
Specifically, we compare results for the static structure factor $S(k)$ and the radial distribution function $g(r)$ \cite{hansen13} obtained from two different DFT approximations and compare with simulation results in order to identify strengths and shortcomings of the DFT theories.
We identify two types of errors arising in the DFT results for $S(k)$, specifically a small systematic error in determination of peak position and an overestimation of peak height.
However, by comparing with the MC simulations, we can understand the scale of the shift and so compensate for the errors in order to use the DFT results in order to accurately predict the location of the peaks in $S(k)$.
This therefore allows to identify the characteristic wavenumbers $k$ of the density modes favoured by the system.
This analysis also allows to understand how these errors manifest in DFT calculations for the density profiles in the various structured phases.
In looking for hallmarks of structured phases in our MC computer simulations of the liquid phase, in addition to calculating $S(k)$ and $g(r)$, we also calculate bond orientational order parameters and identify how they vary on approaching the structured phase from liquid state side of the phase diagram.

This paper is structured as follows: In Sec.~\ref{sec:system} we briefly explain the pair potential defining our model system and the parameters on which the state of the system depends.
In Sec.~\ref{sec:simulation_theory} we first explain the various MC simulation approaches developed, including both the grand-canonical and Gibbs ensemble MC schemes we use. Then, in Sec.~\ref{subsec:DFT} we describe the two DFTs that we have applied and how to obtain the correlation functions. In Sec.~\ref{sec:results} we present our results, structured as follows:
In Sec.~\ref{subsec:liquidcluster} we consider the liquid-cluster transition, as an example to illustrate the changes in structural properties across a phase transition using GCMC simulations. By direct simulation of the phase coexistence using GEMC, we demonstrate that the dominant density modes in the liquid can be associated with the wavectors shaping the structured phases.
In Sec.~\ref{subsec:liquidstability} we present results from a stability analysis of a HCSS bulk liquid. We determine the relevant specific density modes and corresponding wavevectors introduced by the shoulder part of the pair-interaction, elucidating how the ordering arises from the interplay between these and those from the entropic volume-exclusion effects of the hard cores. The influence of the shoulder-interaction on the freezing transition of a {pure hard-core} fluid is also discussed.
In Sec.~\ref{subsec:liquid} we compare the liquid state static structure factors $S(k)$ obtained using GCMC with the analytic results using DFT. This allows us to highlight shortcomings and capabilities of two different DFTs and quantify errors made in the determination of the main structural properties of the liquid.
In Sec.~\ref{subsec:solids} the effect of the latter errors on the predicted equilibrium density profiles $\rho({\bf r})$ is illustrated by comparison to GCMC results. Further, we demonstrate that the analytic predictions from DFT for the bulk liquid suffice to accurately determine the significant wavevectors shaping striped, clustered and holed phases of the HCSS system.
Finally, in Sec.~\ref{sec:conc} we finish with a few concluding remarks.


\section{System}
\label{sec:system}
In our two dimensional system the particles interact via the spherically symmetric {hard-core, square-shoulder} (HCSS) potential, which can be split up into the hard disk interaction (with diameter $\sigma$) and an adjacent soft, repulsive, square shoulder interaction corona of diameter $\lambda \sigma$ and an interaction strength $\epsilon > 0$. The pair potential $\Phi(r)$ thus reads:
\begin{equation}\label{potential}
	\beta \Phi(r)=\begin{cases}
		\infty  & r\leq \sigma \\
		\beta \epsilon &\sigma < r <\lambda\sigma, \\
		0 &\lambda\sigma \leq r.
	\end{cases}
\end{equation} 
Throughout this work we use the {hard-core} diameter $\sigma$ as the unit length of the system (i.e., $\sigma \equiv 1$). Further, we introduce the temperature $T$ via $\beta = 1/(k_{\rm B} T$), with $k_{\rm B}$ being the Boltzmann constant. The properties of the system depend solely on the dimensionless combination $\beta\epsilon$, so for simplicity we henceforth set $\beta \equiv 1$ and refer to the reciprocal shoulder height as the temperature, i.e., $T = 1/ \epsilon$. We introduce the {average} bulk density $\rho_{\rm b}$ with its dimensionless counter-part $\rho_{\rm b}^\star = \rho_{\rm b} \sigma^2$. Further, we define the packing fraction $\eta=\pi\rho_{\rm b}\sigma^2 / 4 $.


\section{Simulation and theory}
\label{sec:simulation_theory}

We {analyse} our system with both extensive MC based simulations \cite{Allen2017,frenkel_smit} and classical DFT based calculations \cite{Evans1979, hansen13}. The essential conceptual elements of both approaches are {summarised} in the following subsections.

\subsection{Simulation methods}
\label{subsec:simulations}

\subsubsection{MC simulations in the grand-canonical ensemble}
\label{subsubsec:GCMC}

We have performed {large-scale} MC simulations in the grand-canonical ensemble (GCMC), imposing particular values for the temperature $T$, the volume $V$ and the chemical potential $\mu$ of the system \cite{Allen2017, frenkel_smit}. The number of particles $N$ can vary within the system but has been limited -- for practical reasons -- to 20~000 particles. The ensemble is confined within a square box for the results presented in Sec.~\ref{subsec:liquidcluster} -- \ref{subsec:liquid}, whereas we use rectangular boxes with aspect ratio $1:\sqrt{3}$ for the solid state simulations presented in Sec.~\ref{subsec:solids}, in order to {minimise} frustration from straining the crystal via the applied periodic boundary conditions. Three types of MC ``moves'' are implemented in the course of a GCMC simulations: translational moves, particle insertions, and particle deletions, with respective probabilities $\alpha_{\rm t}$, $\alpha_{\rm i}$, and $\alpha_{\rm d}$ and suitably adapted acceptance-/rejection-criteria \cite{Allen2017,frenkel_smit}. In our simulations we have assumed that $\alpha_{\rm t} = \alpha_{\rm i} = \alpha_{\rm d} = 1/3$ and have applied the three types of moves in a random consecutive manner. Typical simulations extend in total over $15-20 \times 10^9$ of such MC-moves. 
Equilibration of the energy was typically observed after $4-5 \times 10^9$ MC-moves (see Appendix~\ref{A1}) and ensemble averages were performed using 150-250 states drawn from the end of the simulation separated by $4 \times 10^8$ MC-moves.

\subsubsection{Gibbs ensemble MC simulations}
\label{subsubsec:GEMC}

The Gibbs ensemble represents a very particular ensemble within the framework of statistical mechanics \cite{Allen2017,frenkel_smit, Panagiotopoulos1987}. The entire system is considered in the canonical (i.e., {$NVT$}) ensemble, which is subdivided into two subsystems (confined in sub-boxes) which are assumed to be in phase coexistence, meaning that the two subsystems are at equal temperature $T$, equal pressure $P$, and equal chemical potential $\mu$. This is achieved by introducing three types of MC ``moves'': translational moves of particles inside each of the boxes, particle exchange between the two boxes and volume exchange between the two subsystems while maintaining the total volume of the system. Related acceptance/rejection criteria for each type of moves can be found in the literature \cite{Allen2017,frenkel_smit}. Periodic boundary conditions are applied to each sub-box.

In our simulation we start from two square sub-boxes (indices `1' and `2') with $N_1 = N_2 =1500$ particles, the total number of particles, $N_1+N_2 = 3000$ being fixed. The initial box size is chosen such that $V_1=V_2=V/2$, where the total volume $V=N/\rho_{\rm b}$, where $\rho_{\rm b}$ is the starting total average density and $N=N_1+N_2$. The positions of the particles are {initialised} in hexagonal lattices. 

The simulations are performed via consecutive blocks where each block represents a sequence of the following consecutive steps: (i) $n_{\rm trans}=200$ attempts for translational MC-moves of randomly chosen particles, i.e., 100 moves per sub-box; (ii) a single $n_{\rm volume}=1$ volume change, changing the volume of one sub-box by $\pm\Delta V$ at the cost of the other, where $\Delta V$ is drawn uniformly from the range $\Delta V\in[0,4]$, and (iii) $n_{\rm swap}=500$ attempts to swap particles from one sub-box to the other, where in each step the box in which a particle is deleted is selected randomly.
The position of a newly created particle is chosen uniformly from within the hosting sub-box; the move is rejected if an overlap of hard cores with the existing particles occurs. The simulation is essentially a sequence of repeated blocks, being continued until a total number of MC-move attempts $n_{\rm moves}=n_{\rm blocks}\times(n_{\rm trans}+n_{\rm volume}+n_{\rm swap})\geq8\times10^9$ is attained. For the ensemble averages (e.g., to calculate the structure factor -- see below) each box is treated as a $\mu VT$-ensemble of its own, analogous to the GCMC calculations. 

\subsection{DFT for the HCSS system}
\label{subsec:DFT}

\subsubsection{DFT formalism and thermodynamics}
\label{subsubsec:DFT_thermodynamics}

The thermodynamic and structural properties of our system can alternatively be calculated within the framework of classical DFT. This approach is based on the grand potential functional $\Omega [\rho]$, which in the absence of any external potentials reads \cite{Evans1979, hansen13}:
\begin{equation}\label{DFT2}
\Omega\left[ \rho \right] = {\cal F}[\rho] - \mu\int \rho({\bf r}) d {\bf r}.
\end{equation} 
$\Omega[\rho]$ is a functional of the {one-body} density profile, $\rho({\bf r})$, and is formed from a Legendre transform of the Helmholtz free energy functional ${\cal F}[\rho]$, where $\mu$ is the chemical potential of the system.

The key features of the functional $\Omega[\rho]$ are that (i) $\Omega[\rho]$ is {minimised} by the equilibrium one particle density profile of a system, $\rho_{\rm eq}({\bf r})$, and (ii) for $\rho_{\rm eq}({\bf r})$, the functional $\Omega[\rho]$ takes the value of the thermodynamic grand potential of the system, i.e., $\Omega = \Omega[\rho_{\rm eq}]$.
In a homogeneous liquid, the equilibrium density profile $\rho({\bf r})$ (henceforward, we drop the subscript `eq') is equal to the bulk density $\rho_{\rm b}$. However, in the cluster, stripe, crystal and other such phases the equilibrium density profile $\rho({\bf r})$ is nonuniform.

The Helmholtz free energy can be split into two contributions:
\begin{equation}\label{DFT2_a}
{\cal F}[\rho] = {\cal F}_{\rm id}[\rho]+{\cal F}_{\rm ex}[\rho].
\end{equation}
The first term is the (exact) ideal gas contribution, which can be written as \cite{Evans1979, hansen13}
\begin{equation}\label{DFT3}
{\cal F}_{\rm id}[\rho] = k_{\rm B} T \int \rho({\bf r}) \left[ \log
	\Lambda^2 \rho ({\bf r}) - 1 \right] d {\bf r},
\end{equation}
where $\Lambda$ is the thermal de Broglie wavelength. 
The second term in Eq.~\eqref{DFT2_a} is the excess Helmholtz free energy functional ${\cal F}_{\rm ex}[\rho]$, which incorporates all contributions beyond that of the ideal gas, due to the particle interactions. For the overwhelming majority of systems, an expression for ${\cal F}_{\rm ex}[\rho]$ can only be given in an approximate manner. To do this, we split the particle pair interaction potential in Eq.~\eqref{potential}, into a hard core (index `hc') and a square shoulder (index `ss') part:
{
\begin{equation}\label{DFT_HC}
\Phi_{\textrm{\rm hc}}(r)=\begin{cases}
	\infty  & r\leq 1 \\ 
	0 & 1 < r\\ 
\end{cases}
\end{equation}
\begin{equation}\label{DFT_SS}
	\Phi_{\rm ss}(r)=\begin{cases}
	\epsilon  & r\leq \lambda\\ 
	0 &\lambda < r.\\ 
\end{cases}
\end{equation}
}
Thus, the full potential $\Phi(r)=\Phi_{\rm hc}(r)+\Phi_{\rm ss}(r)$.
Correspondingly, we can split ${\cal F}_{\rm ex}[\rho]$ into a contribution due to the hard core and a remainder, associated with the shoulder:
\begin{equation}\label{DFT_RPA}
	{\cal F}_{\rm ex}[\rho] = {\cal F}_{\rm ex}^{\rm hc}[\rho]+
	{\cal F}_{\rm ex}^{\rm ss}[\rho] .
\end{equation}
Further, we assume that we can treat (in the sense of the random phase approximation -- RPA \cite{hansen13}) the contribution due to the square shoulder potential as a perturbation to the hard core.
The resulting RPA approximation for ${\cal F}_{\rm ex}^{\rm ss}[\rho]$ reads \cite{hansen13}
\begin{equation}\label{DFT_RPA_fex}
{\cal F}^{\rm ss,RPA}_{\rm ex}[\rho]=\frac{1}{2}\int\int\rho({\bf r})\rho({\bf r'})\Phi_{\rm ss}(|{\bf r}-{\bf r'}|) d{\bf r} d{\bf r'}.
\end{equation}
Note that via Eqs.~\eqref{DFT_SS} and \eqref{DFT_RPA_fex} we have followed a standard practice of extending the shoulder of the potential `inside' the hard core. This mean-field approximation is justified as long as $\Phi_{\rm ss}(r)$ is fairly long ranged and the shoulder height $\epsilon$ in Eq.~\eqref{DFT_SS} is reasonably small. However, in practice the RPA approximation can often do better than one might expect, even when these conditions are only loosely met \cite{hansen13, Archer2008, Archer2017}.

For the contribution to the free energy ${\cal F}_{\rm ex}^{\rm hc}[\rho]$, which originates from the {hard-core} interactions, we consider here two different approximations: (i) a very simple approximation, namely a local density approximation (LDA) \cite{hansen13} and (ii) a fundamental measure theory (FMT) based functional \cite{Roth2010,Roth2012}. Both functionals are briefly be introduced in the following.

The LDA functional assumes that the density profile varies sufficiently slowly and that at every point ${\bf r}$, where the density is $\rho({\bf r})$, we can approximate the local contribution to the free energy by that of a bulk liquid with corresponding bulk density $\rho$. This requires the corresponding bulk hard particle fluid equation-of-state (EOS); here we use the one obtained within scaled particle theory (SPT) \cite{hansen13, Helfand1961, Reiss1959}. For the two-dimensional hard disk fluid, the LDA functional together with the SPT EOS is
\begin{align}\label{SPT6.2}
{\cal F}^{\rm LDA}_{\rm ex}\left[\rho\right] = 
k_{\rm B}T\int  \rho ({\bf r}) \left[( - 1 - \log{(1-\eta({\bf r}))} + (1 - \eta({\bf r}))^{-1} \right] d{\bf r},
\end{align}
where the local packing fraction {$\eta({\bf r}) = \pi \rho({\bf r}) / 4$}. 
Experience shows that the above functional fails when the density profile $\rho(\mathbf{r})$ varies on length scales close to that of the {core diameter}, e.g.\ for fluids in the vicinity of hard walls or for crystalline states \cite{evans92}, but for clusters and aggregations of multiple particles, we can expect it to be qualitatively reliable \cite{Glaser2007,Archer2008,Archer2013b}. The results presented below show well when the LDA approximation is sufficient and when it is not, in which case one must use the more sophisticated FMT discussed next.

The FMT functional for hard disks \cite{Roth2012} has been shown to successfully describe the entropic freezing of pure hard discs into a hexagonal crystal, with densely packed arrangement of these particles, having density profiles $\rho({\bf r})$ with sharp Gaussian-like peaks at the appropriate particle positions in the hexagonal lattice.
{In sharp contrast, unlike the FMT, the LDA is unable to predict the freezing of the pure hard-disk (or hard-sphere) fluid.}
The FMT formalism uses so-called weight functions $\omega_\alpha ({\bf r})$ and $\omega^{(m)}({\bf r})$, which by convolution with the one-particle density $\rho({\bf r})$ lead to weighted densities, that are either scalar 
\begin{equation}
n_\alpha ({\bf r}) = 
\left[ \rho\otimes\omega_\alpha \right] ({\bf r}), \hspace{1cm} \alpha= 0, 2.
\end{equation}
or tensorial in nature:
\begin{equation}
{\bf n}^{(m)}({\bf r}) = 
\left[ \rho\otimes{\bf \omega}^{(m)} \right] ({\bf r}), \hspace{1cm} m=0, 1, 2 .
\end{equation}
In the above relations the symbol `$\otimes$' represents spatial convolutions in two dimensions; the weight functions $\omega_\alpha (r)$ and ${\bf \omega}^{(m)}({\bf r})$ are specified below. 

One can then use these weighted densities to generate an approximation for the hard core contribution to the free energy ${\cal F}_{\rm ex}^{\rm hc}[\rho]$ as an integral over a function of these weighted densities \cite{Roth2012}:
\begin{multline}\label{FMT_ex}
F^{\rm FMT}_{\rm ex} [\rho] = k_{\rm B}T \int \Psi({\bf r}) d{\bf r}
= k_{\rm B }T \int \Bigg[ -n_0 ({\bf r}) \log{(1-n_2 ({\bf r}))}\\
+\frac{1}{4\pi(1-n_2 ({\bf r}))} \left( \frac{19}{12}({\bf n}^{(0)}({\bf r}))^2-\frac{5}{12}{\bf n}^{(1)}({\bf r})\cdot {\bf n}^{(1)}({\bf r})-\frac{7}{6}{\bf n}^{(2)}({\bf r}) \cdot {\bf n}^{(2)}({\bf r})\right) \Bigg] d{\bf r}.
\end{multline}
The integrand of the above relation, i.e., the excess free energy density $k_{\rm B} T\Psi({\bf r})$, is obtained from an ansatz which is motivated by dimensional analysis, assuming it must have dimension {length$^{-2}$} and so only has combinations of weight functions consistent with this. This is then combined with a set of differential equations for the bulk pressure, arising from SPT \cite{Roth2010,Roth2012}. The free coefficients in the solutions to these differential equations are determined by requiring that the resulting Helmholtz free energy yields the SPT EOS of for hard disks with bulk density $\rho$ and is well behaved in certain limiting cases of extreme confinement. The resulting Eq.~\eqref{FMT_ex} is valid over a broad range of densities up to and beyond freezing and recovers the exact {low-density} limit.
 
The scalar weight functions introduced above are \cite{Roth2012}
\begin{equation}
\omega_0(r)=\frac{\delta(R-r)}{2\pi R} ~~~~ {\rm and} ~~~~~
\omega_2(r)=\Theta(R-r),
\end{equation}
where {$R=\sigma/2=1/2$}, $\delta(r)$ is the Dirac delta-function and $\Theta(r)$ is the Heaviside step-function. The tensorial weight functions are given by 
\begin{equation}
{\bf \omega}^{(m)}({\bf r})=\delta(R-|{\bf r}|)\underbrace{{\bf \hat{r}}\dots{\bf \hat{r}}}_{\text{$m$-times}},
\end{equation}
where the tensor-rank $m$ of the weight function arises from $(m-1)$ tensor products of the unit vector ${\bf \hat{r}}$ with itself.

{To calculate the equilibrium density profile $\rho({\bf r})$, we use numerical {minimisation} of the DFT functionals via the iterative Picard algorithm described e.g.\ in Refs.~\cite{Roth2010, hughes2014introduction}. Further details are also given in the Appendix.}
Since both the LDA-RPA [Eqs.~\eqref{SPT6.2} and \eqref{DFT_RPA_fex}] and the FMT-RPA [Eqs.~\eqref{FMT_ex} and \eqref{DFT_RPA_fex}] are constructed so that they reproduce the SPT equation of state in the uniform density limit, for both DFTs the chemical potential for a uniform bulk liquid is 
\begin{equation}\label{LS_mu}
\mu=\log(\Lambda^2\rho)+\frac{\eta(3-2\eta)}{(1-\eta)^2}-\log{(1-\eta)}+4\eta\epsilon\lambda^2.
\end{equation}

\subsubsection{Structure}
\label{subsubsec:structure}

Classical DFT not only provides thermodynamic properties, it also gives structural {(pair correlation)} information about the system. One route to this is via the density profiles{, typically using the test-particle method, where a single particle is fixed, treating it as an external potential. The DFT is then minimised to obtain the density profile of the surrounding fluid in the presence of this fixed particle. The resulting profile then directly yields the radial distribution function -- see e.g.\ Ref.~\cite{archer2003solvent}, for further background on this approach. Here, we follow another approach, going} via the direct correlation functions. To be more specific, the two-particle direct correlation function $c^{(2)}({\bf r},{\bf r}')$ for an inhomogeneous fluid can be readily obtained as the following functional derivative of the excess Helmholtz free energy \cite{hansen13}:
{
\begin{equation}\label{eq2.1}
	c^{(2)}({\bf r},{\bf r'}) = -
 \frac{\delta^2{\cal F}_{\rm ex}[\rho]}{\delta\rho({\bf r})\delta\rho({\bf r'})}.
\end{equation}
}
This function {characterises} the two-point density correlations in the system, due to the interactions between the particles. In the homogeneous bulk liquid where $\rho({\bf r})=\rho_{\rm b}$, this correlation function becomes isotropic, i.e. $c^{(2)}({\bf r},{\bf r}')=c^{(2)}(r=|{\bf r}-{\bf r}'|)=c(r)$. The static structure factor, $S(k)$, of such a homogeneous liquid is related to the Fourier transform (FT, indicated by a hat) of the direct pair correlation function as follows
\begin{equation}\label{DFT14}
S(k)=\frac{1}{1-\rho_{\rm b}\hat{c}(k)}.
\end{equation}
The static structure factor $S(k)$ is in turn related to the radial distribution function $g(r)$ as follows \cite{hansen13}:
\begin{equation}\label{ST2}
g(r)=1+(2\pi\rho)^{-1}\int_0^\infty(S(k)-1)kJ_0(kr)dk. 
\end{equation}
Due to the dimensionality of the problem at hand, the transformation from $k$- to $r$-space is {realised} via the inverse Hankel transform, with $J_0(x)$ being the Bessel function of the first kind of order 0. 

Since the Helmholtz free energy functional has been split in an additive manner into the {hard-core and the square-shoulder} contribution -- see Eq.~\eqref{DFT_RPA} -- a related splitting of the direct correlation function (both in $r$-and $k$-space) can be done by virtue of Eq.~\eqref{eq2.1} as follows
{
\begin{equation}
\hat{c}(k) = \hat{c}_{\rm hc}(k) + \hat{c}_{\rm ss}(k) = 
 \hat{c}_{\rm hc}(k) - 
 \hat{\Phi}_{\rm ss}(k).
\end{equation}
}
The simplicity of the last term arises due to the quadratic nature of ${\cal F}_{\rm ex}^{\rm RPA}[\rho]$ -- see Eq.~\eqref{DFT_RPA_fex}. 
The required Fourier transform of the shoulder potential is
\begin{equation} \label{phi:ss_k}
\hat{\Phi}_{\rm ss}(k) = 8\pi\epsilon\lambda\frac{J_1(\lambda k)}{k} ,
\end{equation}
where $J_1(x)$ is the Bessel function of first kind of order 1.

With the LDA- and the FMT-functionals at hand -- see Eqs.~\eqref{SPT6.2} and \eqref{FMT_ex}, respectively -- one can calculate in a straightforward manner (albeit somewhat lengthy in the case of the FMT) the function $\hat{c}_{\rm hc}(k)$.

Using Eqs.~\eqref{eq2.1} and \eqref{SPT6.2} one finds in the case of the LDA the constant value
\begin{equation}
\hat{c}^{\rm LDA}_{\rm hc}(k) = \frac{\eta[4+\eta(\eta-3)]}{\rho(\eta-1)^3}.
\end{equation}
In the case of the FMT functional the situation is more involved, where the direct correlation function is obtained in terms of derivatives of the excess Helmholtz free energy density $\Psi({\bf r})$ with respect to the weighted densities -- see Eq. \eqref{FMT_ex}. The subsequent Fourier transform of the weight functions leads to \cite{Thorneywork2018}:
\begin{equation}\label{FMT_ckp}
\hat{c}^{\rm FMT}_{\rm hc}(k) = -\sum_{\alpha,\gamma}\frac{\partial^2\Psi}{\partial n_\alpha \partial n_\gamma} \hat{\omega}_\alpha(-k) \hat{\omega}_\gamma(k), 
\end{equation}
{by which}, together with Eq.~\eqref{FMT_ex}, one eventually obtains
{
\begin{multline} \label{c_FMT_hc}
\hat{c}^{\rm FMT}_{\rm hc}(k)=\frac{\pi}{6(1-\eta)^3k^2}\bigg[-\frac{5}{4}(1-\eta)^2k^2J_0(k/2)^2\\+\left( 4((\eta-20)\eta+7)+\frac{5}{4}(1-\eta)^2k^2\right)J_1(k/2)^2\\+2(\eta-13)(1-\eta)k J_1(k/2)J_0(k/2)\bigg].
\end{multline}
}
Since the static structure factor $S(k)$ can easily be calculated from $\hat c(k)$ via Eq.~\eqref{DFT14} and since in both {approximations} $\hat c(k)$ is given by analytic expressions, one can -- in principle -- write down the expression for $S(k)$.

In the LDA this is straightforward, giving the following closed expression
{
\begin{equation}\label{S_LDA}
S_{\rm LDA}(k) = \left(\frac{1+\eta}{(1-\eta)^3}+8\eta
\epsilon\lambda\frac{J_1(\lambda k)}{k}\right)^{-1}.
\end{equation}
}
A related expression is found for $S_{\rm FMT}(k)$ which is considerably more lengthy (in view of the complexity of $\hat{c}^{\rm FMT}_{\rm hc}(k)$, Eq.~\eqref{c_FMT_hc}) and so is not reproduced here. 

\subsubsection{Stability analysis}
\label{subsubsec:stability}

Identification of the regions of the phase diagram where the uniform bulk liquid becomes unstable simultaneously enables identification of the regions where the formation of ordered phases can be expected. In the regions where the bulk liquid is unstable, it is because it becomes unstable against the growth of periodic modulations of the density, {characterised} by a specific wavevector. Here, we demonstrate that the specific wavevectors that play a significant role in describing the onset of instability of the liquid are also sufficient to predict the length-scales that one can expect in (possibly highly complex) ordered structures.

When considering the stability of a liquid, one can take either a thermodynamic/structural or a dynamical perspective. Within the DFT framework, these are closely related, since both originate from the free energy functional. A dynamical analysis is based on dynamical density functional theory \cite{marconi1999dynamic, Archer2004, hansen13, te2020classical}. To identify the limit of linear stability of the bulk liquid, one approach is to identify the locus in the $(T, \rho)$-plane where the first peak of the static structure factor (located at $k_{\rm c}>0$) -- see \eqref{DFT14} -- is diverging, i.e., $S(k_{\rm c}) \to \infty$.
This marginal stability limit of the liquid against the critical wavenumber $k_{\rm c}>0$, is referred to here as the $\lambda$-line, following the terminology of Ref.~\cite{archer2007model}. This should not to be confused with our use of $\lambda$ as the parameter defining the range of the pair potential \eqref{potential}.
From the dynamical perspective, inside the $\lambda$-line the uniform liquid is dynamically unstable against periodic density modulations with this unstable wavenumber, where (at least within the linearised regime) these modes grow over time without any (free) energetic barrier to surmount \cite{archer2012solidification}. Either way, one can formulate the necessary conditions for locating the $\lambda$-line based on properties of $S(k)^{-1}$ [{see} Eq.~\eqref{DFT14}] via the following criteria: (i) the extremum condition, i.e., $\frac{d}{dk}S(k)^{-1}\big|_{k = k_{\rm c}}=0$ and (ii) the divergence condition, i.e., $S(k_{\rm c})^{-1} \to 0$. 

By virtue of the simplicity of the LDA expression for the static structure factor [see Eq.~\eqref{S_LDA}] the extremum condition can be formulated in a closed expression

\begin{equation}\label{SPT12}
k_{\rm c}=\frac{j_{2,1}}{\lambda},
\end{equation}
where $j_{i,n}$ is the $n^{\rm th}$ zero of the Bessel function of the first kind of $i^{\rm th}$ order.
Since $k_{\rm c}$ is evaluated via an extremum condition it is worth mentioning that for the LDA-RPA not only is the position of the main peak of $S(k)$ given by the zeros of this specific Bessel function, but also all the other maxima and minima.
Consequently, the number of critical wavevectors that have to be considered in a stability analysis strongly depend on the interaction range $\lambda$, since for larger $\lambda$, more local minima of $S(k)^{-1}$ arise in the relevant interval {$0<k\lesssim 2\pi$}.
Here, we focus on the instability of the bulk liquid against the single wavenumber $k_{\rm c}$, corresponding to the global minimum of $S(k)^{-1}$. However, we should mention that one can tune the system so that there are two (or more) critical wavenumbers, a topic we will elaborate elsewhere. Using the divergence conditions and inserting the critical wavenumber $k_{\rm c}$ into Eq.~\eqref{S_LDA}, we obtain the following parametric curve for the $\lambda$-line:

\begin{equation}\label{LS1}
\frac{1}{\epsilon}=\frac{8\eta(\eta-1)^3}{1+\eta}\frac{J_1(j_{2,1})}{j_{2,1}}\lambda^2.
\end{equation}
The corresponding calculation for the FMT-RPA is not analytically tractable, but using numerical methods the critical wavenumber $k_{\rm c}$ and the $\lambda$-line can be obtained for the FMT-RPA, by following the general approach outlined above. Discussion and comparison of the LDA-RPA with the FMT-RPA results is made in Sec.~\ref{subsubsec:stability}.

\section{Results}\label{sec:results}

\subsection{Structural properties across the liquid-cluster phase transition}\label{subsec:liquidcluster}

To identify signatures of incipient self-assembled structures in the liquid phase and to elucidate the changes in the structural properties as the liquid becomes unstable to periodic phases, we first focus on the liquid-cluster phase transition. This has been well studied for the different but somewhat related `short-range-attraction, longer-range-repulsion' (SALR) system \cite{Imperio2004}.
We find that at low densities, the phase ordering of the SALR and HCSS systems are similar, but at higher densities they can be different, depending on the value of $\lambda$. The low-temperature clustered phase is {characterised} by two distinct length-scales, which shape the self-assembled pattern:
(i) The larger one is the typical cluster spacing, $a_{\rm c}=2\pi/k_{\rm c}$, which governs both the area occupied by a single cluster as well as the self-assembled hexagonal ordering pattern of the clusters. Below, we show that $a_{\rm c}$ is related to the wavevector $k_1$, corresponding to the first peak in $S(k)$. (ii) The shorter length-scale is the length-scale of freezing of particles within clusters, $a_{\rm f}=2\pi/k_{\rm f}$ (where $k_{\rm f}$ is the freezing wavevector), governing the spacing and the arrangement of individual particles within a given cluster.
Below, we show that $a_{\rm f}$ is related to the wavevector of one of the higher-$k$ maxima in $S(k)$. For example, for the case where $\lambda=3.7$ considered below, this corresponds to the fourth maximum, at $k_4$. While $a_{\rm c}$ is predominantly determined by the shoulder width $\lambda$, $a_{\rm f}$ is in general associated with the volume-exclusion effects of the hard core. In fact, as the density increases, $k_{\rm f}$ converges towards $k_{\rm hex} = 4 \pi/\sqrt{3} \approx 7.26$ in the limit of a closest hexagonal packing of hard cores, occurring at $\eta \approx 0.906$.

In order to obtain deeper insight into the implications of the DFT-based results in Sec.~\ref{subsubsec:stability} and to determine the relevance of the wavevectors corresponding to peaks in $S(k)$ on the structure, we first need to verify that these are indeed the wavevectors governing the density fluctuations in a marginally stable liquid and also are the ones that ultimately shape the structures that emerge as the liquid becomes unstable.
We {emphasise} that the investigations discussed here are entirely based on GCMC simulations.
For this purpose we focus on one system, {characterised} by $\lambda = 3.7$. This choice is in some sense arbitrary (as we expect similar results for other such intermediate values of $\lambda$) and is justified by the fact that for shoulder widths around this value, the relevant wavevectors can be nicely displayed in graphical representations.
Further, we fix the values of the chemical potential $\mu$ (as specified below) and the temperature to be located in the vicinity of the liquid-cluster phase transition {-- see the Appendix}.

\begin{figure}[t!]
\centering
\includegraphics[width=0.33\textwidth]{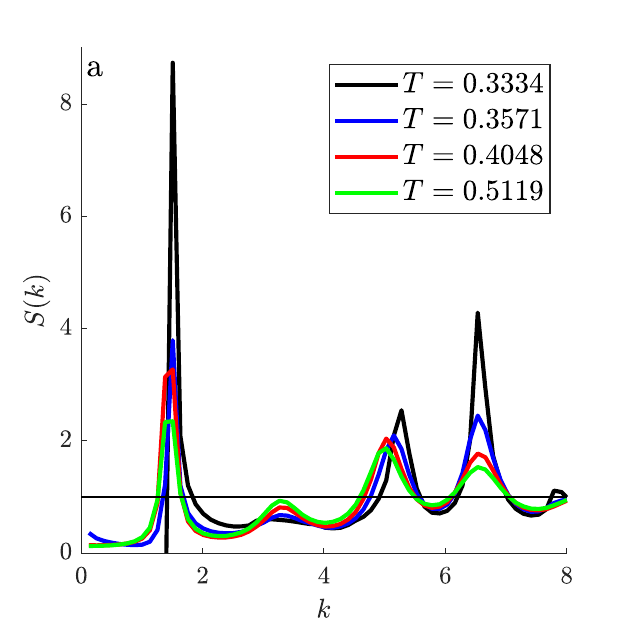}\includegraphics[width=0.33\textwidth]{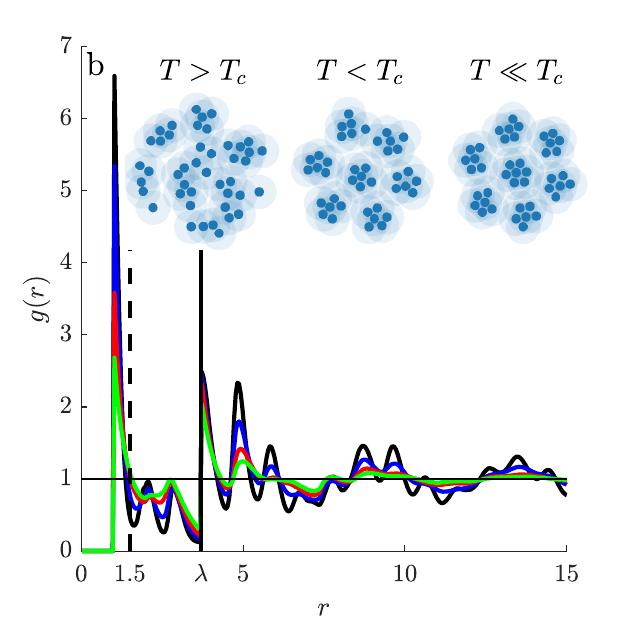}\includegraphics[width=0.33\textwidth]{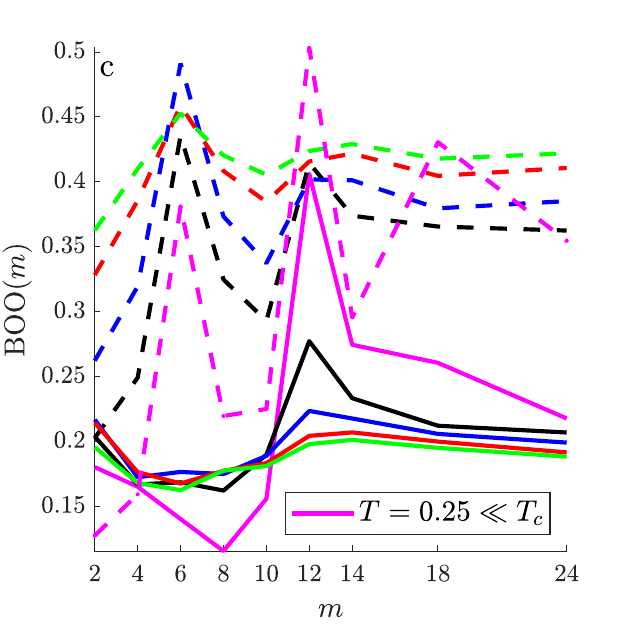}
\caption{
{Panel a) shows the} radially symmetric structure factor $S(k)$ as a function of the wavevector $k$ for a HCSS system with $\lambda$ = 3.7, evaluated for different temperatures $T$ (as labeled) and for constant chemical potential $\mu(\eta =0.2)$; see Eq.~\eqref{LS_mu}. $S(k)$ is obtained via Eq.~\eqref{ST2} from the radial distribution function $g(r)$ (see panel {b}), obtained in GCMC simulations.
{Panel b) shows the} radial distribution function $g(r)$ as a function of $r$, obtained in GCMC simulations for {the different temperatures indicated in the key in panel a)}. At the top of {panel b)} are displayed three characteristic snapshots of the liquid and of the cluster phase. The dashed vertical line indicates the distance $r = 1.5$, relevant to identify nearest neighbours of a tagged particle while the solid vertical line indicates the range of the shoulder $\lambda= 3.7$.
{Panel c) displays the} bond orientational order parameter BOO($m$) for a selection of $m$-values {(those indicated on the axis)} and for different temperatures (as labeled in panel {a}). The data for an additional set of BOO($m$) have been added, which were evaluated for $T = 0.25$ (shown in magenta). {The dashed lines are BOO($m$) evaluated for just nearest neighbours (with distances $r \le 1.5$), while the solid lines are for all} neighbours separated by distances up to the shoulder width, i.e.\ for $r \le \lambda$.}
\label{Fig1}
\end{figure}

\begin{figure}[t!]
\centering
\includegraphics[width=0.66\textwidth]{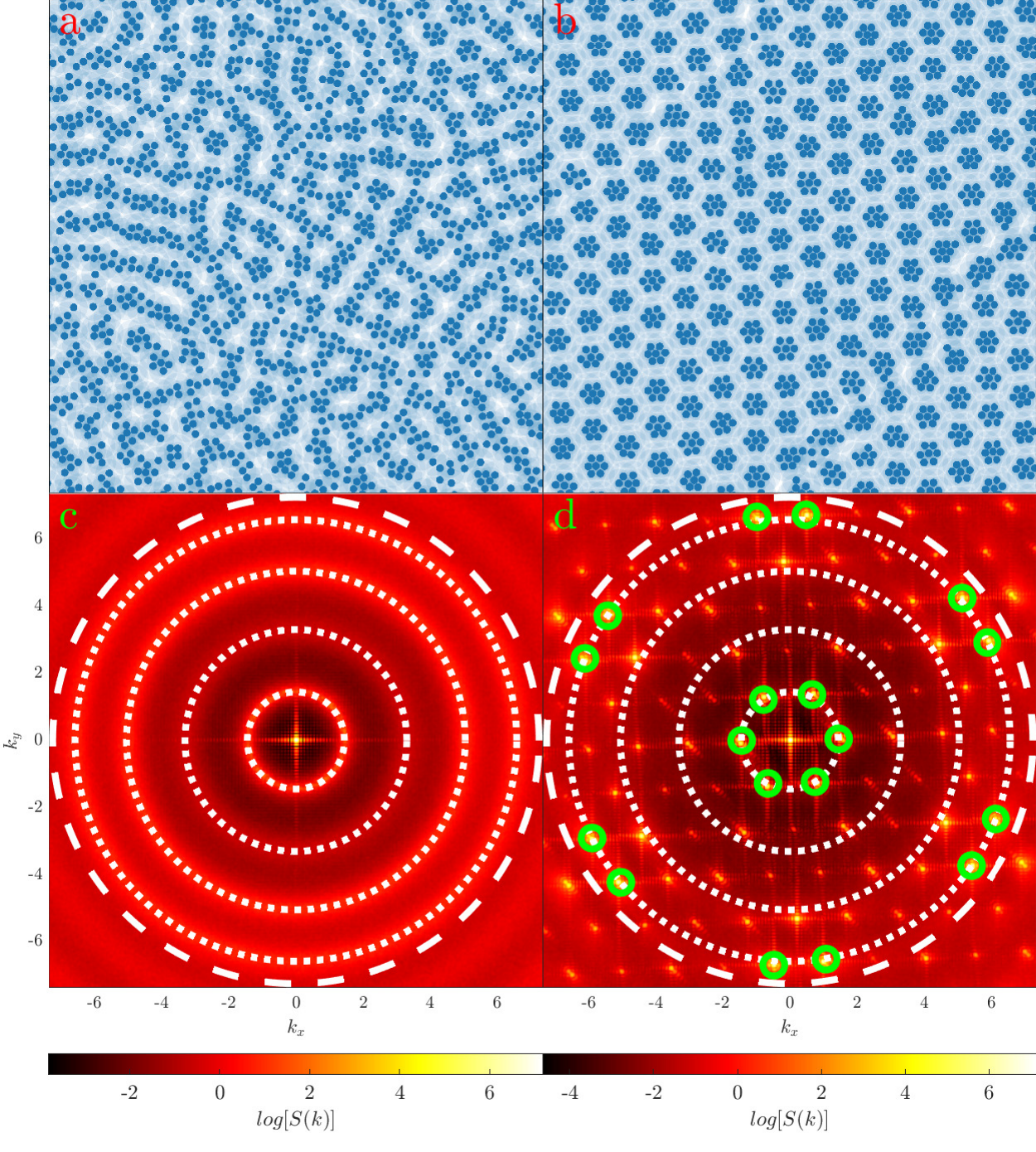}
\caption{{Panels a) and b) show GEMC snapshots of a HCSS system with $\lambda = 3.7$} and $T= 0.3663$; these states are at the transition from the liquid {(a)} to the cluster phase {(b), having average densities $\langle\rho_{\rm liq}\rangle=0.0902$ and $\langle\rho_{\rm cl}\rangle=0.0898$, respectively}.
{Panels c) and d) display} heatmap plots (see colour bars) of the static structure factor $S({\bf k})$ as a function of ${\bf k} = (k_x, k_y)$. The radii of the dotted white concentric circles in both {c) and d)} are the wavectors $k_1, \dots, k_4$ of the radially averaged structure factor $S(k)$ in the liquid phase, corresponding thus to the four main peaks in $S(k)$ in the liquid phase. The radius of the dashed white circle is $k_{\rm hex} = 4 \pi/\sqrt{3}$ (see text). The small {green circles in d)} highlight the Bragg-peaks associated with the hexagonal patterns observed between and within clusters (see text).}
\label{Fig2}
\end{figure}

In Fig.~\ref{Fig1} we present results obtained from GCMC simulations. These are all state points where the DFT predicts the uniform liquid to be unstable, i.e.\ are inside the $\lambda$-line.
In {Fig.~\ref{Fig1}a} we display the static structure factor $S(k)$ at different temperatures, for states located in the vicinity of the liquid-cluster transition and for the fixed chemical potential value $\mu(\eta=0.2)$, as obtained from Eq.~\eqref{LS_mu}.
For this value of $\mu$, the liquid-cluster transition occurs at $T_{\rm c} \approx 0.36$. This is also identified by GEMC -- see Fig.~\ref{Fig2} -- but not for exactly the same value of the chemical potential considered here. Additional evidence for this value of the transition temperature at $\mu(\eta=0.2)$ also comes from inspecting the average packing fraction and bond-order-parameter as a function of temperature; further details are given in the Appendix (see especially Fig.~\ref{SupportGCMC}).
$S(k)$ is obtained via the inverse Fourier transform of Eq.~\eqref{ST2} together with the radial distribution function $g(r)$, which is displayed in {Fig.~\ref{Fig1}b} (and discussed further below).
$g(r)$ is calculated in GCMC simulations, using a simulation box of size $L_{x} \times L_{y} = 100 \times 100$.

The plot of $S(k)$ in {Fig.~\ref{Fig1}a} shows four distinct peaks with values $S(k_i) = S_i$, located at wavevectors $k_i$, $i = 1, \dots, 4$. These four are those that are within the displayed physically relevant range of $k$-values, i.e., $0 < k_i \lesssim k_{\rm hex}$ (i.e.\ there are other maxima at larger-$k$ and there can also be a local maximum at $k=0$, that are not of interest here).
In this plot, we observe that the peak heights $S_1$ and $S_4$ increase significantly as the temperature is decreased. It is interesting to enquire if the so-called Hansen-Verlet (HV) criterion (generalised to 2D systems) applies in the present case \cite{hansen1969phase, costigliola2016freezing, broughton1982molecular, wang2011two}. HV observed empirically for the 3D Lennard-Jones (LJ) system that as the main peak of $S(k)$ passes the threshold $S_1\approx2.85$, the LJ system is prone to freeze. This criterion has been found to roughly apply to a wide variety of different systems, albeit for 2D systems the HV threshold is $S_i\approx5$ \cite{broughton1982molecular, wang2011two}.
For the two higher temperatures displayed in Fig.~\ref{Fig1} (i.e., for $T = 0.5119$ and $T = 0.4048$), the HCSS system is still in the liquid phase, which agrees with what the 2D HV criterion suggests.

At the lower temperature $T= 0.3571$, which is close to {the transition temperature obtained from the MC simulations (see also the Appendix),} $T_{\rm c} \approx 0.36$, where the phase transition to the cluster phase (consisting of a hexagonal lattice of clusters) occurs for the imposed value of $\mu$, $S(k)$ exhibits several pronounced peaks.
This is displayed in {Fig.~\ref{Fig1}a}. The first peak (at $k_1$) has height $S_1\approx4$, which is a little below the 2D HV threshold value. Also, the fourth peak $S_4$ is also prominent, but some way short of the HV threshold.
It is possible that this state is located in the two phase coexistence region between the liquid and the cluster phase.

At the lowest temperature $T = 0.3334$, the system is definitely in the cluster phase, having pronounced peaks with large values of both $S_1$ and $S_4$ (with $S_1$ well above the HV threshold).
In this case, $S_3$ assumes a relatively modest value. As a consequence, the length-scale associated with $k_3$ turns out to play only a minor role in the cluster phase.
It is also worth noting that the values of $k_1$ and $k_4$ remain essentially unchanged as we vary the temperature, while one can observe a {notable} temperature-dependent variation of the $k_2$- and $k_3$-values. As discussed below, the large peaks $S_1$ and $S_4$ and the corresponding wavevectors impact on the shape of the radial distribution function.

The corresponding radial distribution functions, $g(r)$, are displayed {in Fig.~\ref{Fig1}b}. These display the characteristic features of $g(r)$ for a HCSS-system, namely the two distinct discontinuities at {$r = 1 \,\,(= \sigma)$ and $r = \lambda$}. These two discontinuities are to be expected and arise essentially as contributions of the core and of the shoulder to the overall pressure in the system. The magnitudes of the discontinuities are related to the bulk pressure of the system \cite{hansen13, Serratos2012}.
Further, $g(r)$ is {characterised} by oscillations both on a long- and a short-range scale, with wavelengths that can be associated with $k_1$ and $k_4$, respectively. For the two higher temperature liquid states, both types of oscillations decay relatively fast. For the temperature near the transition ($T = 0.3571$) the long wavelength oscillations in $g(r)$ are of large amplitude, indicating strong inter-cluster correlations. At the lowest temperature displayed (in the cluster phase), the short wavelength oscillations also decay rather slowly and with large amplitude, indicating significant correlations of particle positions both within and between the clusters.

Thus, in the case of an intermediate or large shoulder width and at low densities, where particles have sufficient space to move around freely inside a cluster, we can identify the following two stages of cluster formation: (i) below $T_{\rm c}$, but still at sufficiently high temperatures (where $S_1$ is large and well above the HV threshold, but $S_4$ is not), clusters populate a hexagonal lattice, while inside each cluster the particles form a disordered, liquid-like structure; (ii) at even lower temperatures (where we observe $S_4$ approaches or exceeds the HV threshold, along with $S_1$), we find that the particles freeze inside the clusters into hexagonal particle arrangements (with close core-contact).
Furthermore, the clusters themselves are oriented in the same direction. 
For {visualisation} of this trend, typical snapshots are shown along the top of {Fig.~\ref{Fig1}b}. For the liquid-cluster phase transition at higher densities -- which we investigate using GEMC simulations below -- the stages (i) and (ii) occur simultaneously, with a large value of $S_1$, above the HV threshold. The particles are constrained to the area of the clusters, which at higher densities immediately leads to freezing due to packing of the cores. Only at lower densities and large $\lambda$-values are clusters of liquid-like droplets observed.

In {Fig.~\ref{Fig1}c} we display the bond orientational parameters BOO$(m)$ for a selected range of $m$-values{.
For 2D systems, these are defined as \cite{Strandburg1984,Strandburg1988}:
\begin{equation}\label{BOO}
{\rm BOO}(m)=\left\langle \left| \frac{1}{N_{\rm N}}\sum_{\bf r} e^{im\theta_{\bf r}}\right|^2 \right\rangle,
\end{equation}
where $N_{\rm N}$ is the number of neighbours for a tagged particle at position ${\bf r}$ and with $\theta_{\bf r}$ being the angle between an arbitrary but fixed axis and the line connecting the particle at hand with its neighbours within a specified range.}
In evaluating BOO($m$), we consider for a tagged particle only those {neighbours} that are (i) separated by distances $r \leq 1.5$ (a distance which corresponds roughly to the position of the first peak in $g(r)${, so these are the nearest-neighbours}) or (ii) {all particles with distances $r \le \lambda$ from the tagged particle} (i.e., up to shoulder contact).
These two distances are marked in {Fig.~\ref{Fig1}c} by a dashed and a solid vertical line, respectively.
{The average $\langle\cdots\rangle$ in Eq.~\eqref{BOO} indicates a statistical average performed over all particles in the system.}
In addition to the four temperatures considered {in Fig.~\ref{Fig1}a-b, in Fig.~\ref{Fig1}c} we have added also the results for an even lower temperature state (i.e., for $T=0.250$) for which the orientations of the particles within each cluster are fully aligned.
In the case of nearest neighbours, the two peaks of the BOO$(m)$ (for $m=6$ and $m=12$) become more pronounced as the temperature drops below $T_{\rm c}$. In contrast, the transition from the liquid to the cluster phase can be associated with an increase in BOO(12) at shoulder contact. Note that due to the range of the shoulder interaction, the averaging procedure in the evaluation of BOO$(m)$ for shoulder contact involves in general 10 to 20 particles.

The above analysis of the (radially symmetric) structure factor $S(k)$ shows that for the system at hand the transition from the liquid into the cluster phase is {characterised} by a significant increase of $S_1$ and $S_4$. In the following we provide evidence that the wavevectors associated with these peaks in $S(k)$ in the liquid state are identical to the wavevectors that are responsible for forming the cluster phase.
To do this, in Fig.~\ref{Fig2} we now analyse in detail the full angular dependent (i.e.\ vector ${\bf k}$-dependent) structure factor $S({\bf k})$ for the system at hand (i.e., for the {cluster phase that is at coexistence with the liquid, which have average densities $\langle\rho_{\rm cl}\rangle=0.0898$ and $\langle\rho_{\rm liq}\rangle=0.0902$, respectively}). $S({\bf k})$ is obtained in GEMC simulations (see the Appendix, Eq.~\eqref{ST4}), directly from the positions of the particles as input.
In {Fig.~\ref{Fig2}a and \ref{Fig2}b} we display typical snapshots of the two phases involved, while {Fig.~\ref{Fig2}c and \ref{Fig2}d} show heatmap plots of $S({\bf k})$ as functions of ${\bf k} = (k_x, k_y)$.
{Fig.~\ref{Fig2}c} shows the typical radially symmetric result for a fluid state, while {Fig.~\ref{Fig2}d} exhibits the typical Bragg-peaks of a crystalline state.
On top of these, we superimpose four white dotted concentric circles, whose radii correspond -- from the {centre} to the periphery -- to the aforementioned wavevectors $k_1, \dots, k_4$ which correspond to the four main peaks ($S_1, \dots, S_4$) of the radially averaged structure factor -- see Fig.~\ref{Fig1}. An additional white dashed circle of radius $k_{\rm hex} = 4 \pi/\sqrt{3}$ is also displayed.
In the case of {Fig.~\ref{Fig2}d}, we observe that: (i) The circles with radii $k_1$ and $k_4$ match very nicely the brightest Bragg-peaks in $S({\bf k})$ (circled in {green}) in the cluster phase. Thus, the peaks associated with hexagonal patterns observed in $S({\bf k})$ are located with high accuracy on the circles corresponding to wavevectors $k_1$ and $k_4$. (ii) The Bragg-peaks near $k_3$ are found to have a somewhat bigger $k$-value in the cluster phase as compared to the liquid phase (i.e.\ these Bragg-peaks do not lie on the $k_3$ circle), an observation which is in nice agreement with the findings from the GCMC simulations discussed in the previous subsection.
With all this in mind, we can conclude that the cluster phase that emerges as the liquid becomes unstable is shaped by those wavevectors (namely $k_1$ and $k_4$) that predominantly govern the density fluctuations in the liquid. This points to the possibility that once identified, one can tune these wavectors, gaining insight into the complex phase behaviour of HCSS-system and be able to {\it design} self-assembled phases using specific combinations of wavectors. We flesh this idea out further in the following sections, but pursue the `design' aspect in a future publication.

In the following subsections we test whether the two different DFTs introduced above in Sec.~\ref{subsec:DFT} are able to predict (in a quantitative manner) the values of the relevant wavectors.

\subsection{DFT-based stability analysis of the HCSS liquid}
\label{subsec:liquidstability}

\begin{figure}[t!]
\centering
\includegraphics[width=0.5\textwidth]{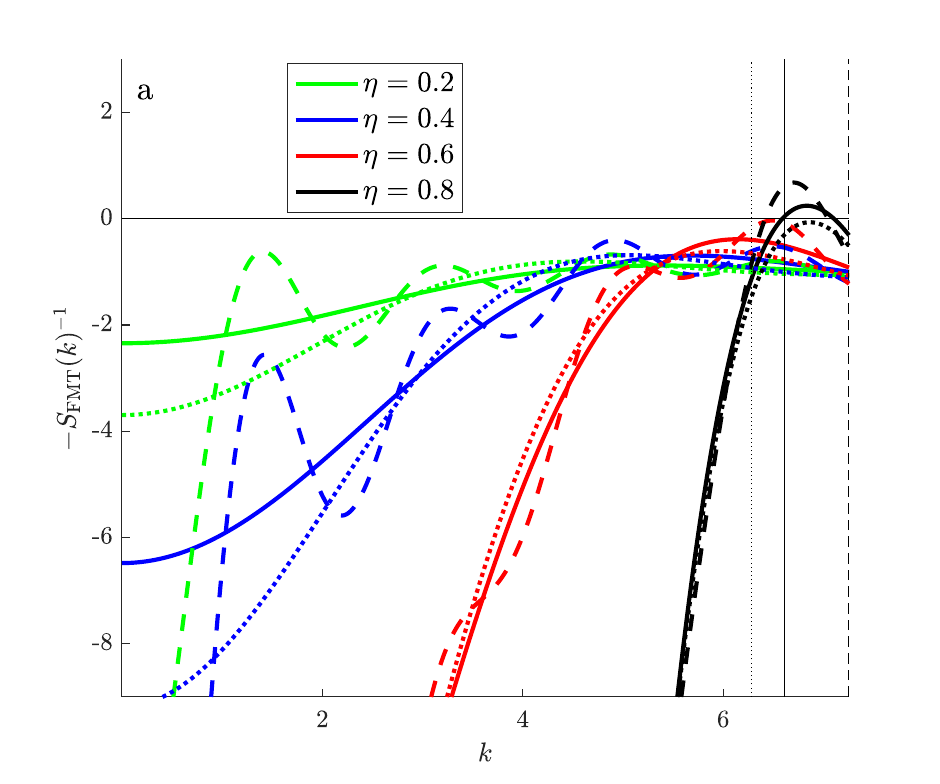}\includegraphics[width=0.5\textwidth]{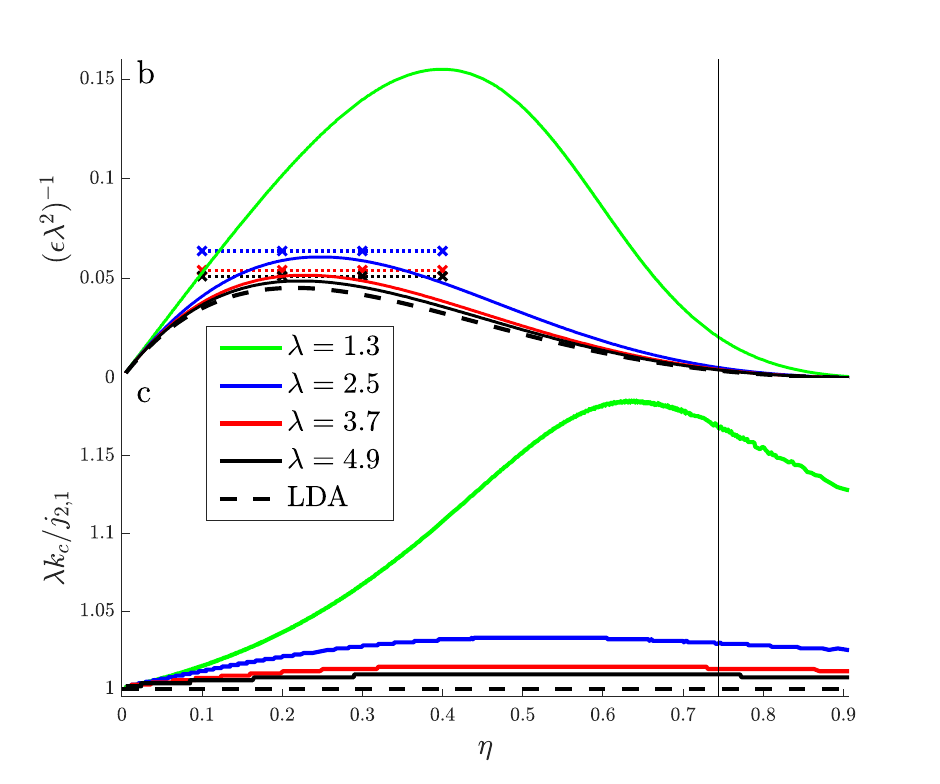}
\caption{{Panel a)}: plots of the negative reciprocal of the structure factor $-1/S_{\rm FMT}(k)$ as a function of $k$,  for different values of the packing fraction $\eta$ (as labeled). The solid lines are for the $\epsilon=0$ pure HC system as obtained via FMT, from Eq.~\eqref{c_FMT_hc}. The dotted and dashed lines are related results for the HCSS-system, with  $\epsilon=1$, and $\lambda=1.3$ (dotted) or $\lambda=3.7$ (dashed), for the same range of values of $\eta$. The black vertical lines indicate the wavenumber for HC contact $k_{\rm con}=2 \pi$ (dotted), the wavenumber associated with closest hexagonal packing $k_{\rm hex} = 4 \pi/\sqrt{3}$ (dashed) and the predicted critical wavenumber for freezing of the pure HC fluid $k_{\rm f}=6.616$ (full).
{Panel b)}: scaled temperature location of the $\lambda$-line as functions of $\eta$, calculated numerically within FMT (solid lines) and within LDA (dashed lines) for different values of $\lambda$ (as labeled). For clarity the curves were multiplied by a factor $\lambda^{-2}$ in order to scale out the shoulder dependency of the LDA results -- see Eq.~\eqref{LS1}.
{Note that when displayed this way, the LDA $\lambda$-line corresponds to the same curve for all values of $\lambda$.}
The horizontal dotted lines (and crosses) indicate the temperatures of the GCMC simulations carried out to investigate specific liquid states (see discussion in Sec.~\ref{subsec:liquid}).
{Panel c)}: critical wavenumbers $k_{\rm c}$ as functions of $\eta$, calculated within FMT (solid lines) and within LDA (dashed lines) for different values of $\lambda$, as labeled.
For clarity the curves were re-scaled by the density-independent, critical wavevector predicted by the LDA {$k_c=j_{2,1}/\lambda$} -- see Eq.~\eqref{SPT12}.
The black solid vertical line {[which extends over both panels b) and c)]} indicates the packing fraction at which entropic freezing is predicted by the FMT for the pure HC fluid $\eta_{\rm f}=0.744$.}
\label{Fig3}
\end{figure}

Having shown which wavectors {characterise} the structure of the liquid and are also involved in the structure of the cluster phase and also shown that monitoring the values of $S_1$ and $S_4$ allows to predict the emergence of ordered structures, we now move on to discuss how DFT fares in predicting these wavectors.
In Fig.~\ref{Fig3} we present results for the negative reciprocal of the structure factor $-1/S(k)$ and for the location in the phase diagram of the $\lambda$-line.
As mentioned already, DFT predicts the liquid to be linearly unstable (i.e.\ within the $\lambda$-line) at temperatures higher than where the cluster formation actually occurs.
This is similar to the case with the SALR class of models, where the connection between the location of the $\lambda$-line and the cluster formation has been investigated in some detail
\cite{Imperio2004, imperio2006microphase, pini2006freezing, archer2007model, archer2007phase, chacko2015two}.
{Figure~\ref{Fig3}a} illustrates how the presence of the square-shoulder interaction modifies the structure factor of the pure HC system. 
Note that we plot $-1/S_{\rm FMT}(k)$, the structure factor from the FMT-DFT, as a function of $k$ rather than $S_{\rm FMT}(k)$ itself, because the exercise of investigating where and how peaks in $S(k)$ {become} large (and within the DFT treatment {diverge} $S(k_i)\to\infty$) is entirely equivalent to investigating where and how peaks in $-1/S_{\rm FMT}(k)$ approach zero from below. 

In {Fig.~\ref{Fig3}a} we compare $-1/S_{\rm FMT}(k)$ for the pure {hard-core} (HC) liquid with that for the HCSS liquid with $T=1$ (i.e.\ $\epsilon=1$) and either $\lambda=1.3$ or $\lambda=3.7$. The pure HC system becomes marginally unstable at the packing fraction $\eta_{\rm f} \approx 0.744$, i.e.\ where $-1/S_{\rm FMT}(k_{\rm f})=0$, with a critical freezing  $k_{\rm f} \approx 6.616$. This freezing is purely entropic nature. At very high densities this peak is located at $k_{\rm hex}=4\pi/\sqrt{3}$, representing the wavenumber that is responsible for a closest hexagonal packing, occurring at $\eta_{\rm hex}=\pi/(2\sqrt{3}) \approx 0.907$.
Note that the packing fraction at which the freezing transition occurs as predicted by {minimisation} of the FMT functional in a 2D domain is at packing fractions somewhat lower than this: The FMT predicts the liquid-solid coexistence packing fractions to be {$\eta_{\rm liq}=0.711$} and {$\eta_{\rm sol}=0.732$} \cite{lin2018phase}, for the liquid and solid respectively. For $0 < \eta <\eta_{\rm f}$ the liquid is linearly stable, but only for {$\eta\leq\eta_{\rm liq}$} is it the global free energy minimum and therefore the thermodynamic equilibrium state.

The location of the peak in $S(k)$ for the high density liquid predicts fairly well the wavenumber determining the length scale of the particle ordering in the pure HC crystal. Since this freezing is entropic in origin (in contrast to the cluster freezing that is driven by the energetics of the shoulder part of the potential), we refer to this wavenumber as the entropically favoured wavenumber in our discussions here.
For the HCSS-liquid (with results shown for $\lambda=1.3$ and $\lambda=3.7$ in Fig.~\ref{Fig3}) we observe a structure factor that shows a similar behaviour to that of the pure HC system for $k\approx k_{\rm f}$. However, at smaller $k$, the smooth curve of $S_{\rm HC}(k)$ is superposed by modulations with maxima occurring at $\approx2\pi n/\lambda$, with integer $n=1,2,3,...$ (a better approximation for the locations of these maxima comes from the locations of the maxima in Eq.~\eqref{phi:ss_k}, i.e.\ Eq.~\eqref{SPT12}).
Each of these peaks correspond to wavenumbers of density modes that may or may not be favoured by the nonlinear couplings in the free energy functional.
At intermediate to high densities the FMT-based HC contribution of the excess free energy functional dominates, leading to a favouring of modes in the vicinity of the peak of the structure factor of the pure HC fluid. However, at lower densities, modes at the other peaks can be favoured. For very high densities (i.e., for $\eta > 0.7$) the modulations in $-1/S_{\rm FMT}(k)$ are no longer visible and only one single peak, that associated with entropic freezing, emerges in the region $k_{\rm con} < k < k_{\rm hex}$, where {$k_{\rm con}=2\pi$}, is the wavevector associated with core contact. Hence -- and as already discussed in the context of Eq.~\eqref{SPT12} -- longer ranged shoulder interactions lead to the introduction of additional peaks in the structure factor and, consequently, other preferred wavevectors that can potentially play a role in the formation of structured phases. We discuss further the role of these additional wavevectors in the liquid phase in Sec.~\ref{subsec:liquid}.

In {Fig.~\ref{Fig3}b} we display the location in the phase diagram of the $\lambda$-line as a function of $\eta$, for {different} values of $\lambda$. By virtue of Eq.~\eqref{LS1}, the $\lambda$-line scales within the LDA-DFT treatment as $T \propto \lambda^2$. Therefore, we display the $\lambda$-line in a scaled form, namely by plotting in the scaled temperature $(\epsilon \lambda^2)^{-1}$ versus $\eta$ plane.
This helps us to compare systems with different shoulder widths $\lambda$ in a more transparent manner.
For small values of $\eta$, the LDA- and FMT-DFT results are in a reasonably good agreement, but rather pronounced differences between the two can be observed at intermediate and high densities. These discrepancies are largest for the smallest $\lambda$-value investigated, while for the larger $\lambda$'s the FMT-based curves essentially converge towards the LDA-DFT results. Eventually, as the shoulder becomes very short ranged, i.e.\ as $\lambda$ approaches unity, the first peak in the static structure factor is roughly located at $k_{\rm c} \approx j_{2,1} \approx 5.135$ -- see Eq.~\eqref{SPT12} -- where it is close to the ($\eta$-dependent) position of the main peak of the HC structure factor at intermediate densities (see Fig.~\ref{Fig3}). This leads to a strong $\eta$-dependence of both the $\lambda$-line and of $k_{\rm c}$ for smaller $\lambda$-values. The predicted freezing packing fraction $\eta_{\rm f}$ of a HC fluid is indicated by the solid vertical line. 

The value of the critical wavector $k_{\rm c}$ related to the $\lambda$-line is plotted as a function of $\eta$ in {Fig.~\ref{Fig3}c}, for various values of $\lambda$. To highlight the differences between the FMT- and the LDA-results for this quantity, in this plot we have divided the FMT $k_{\rm c}$ by the corresponding LDA result, given in Eq.~\eqref{SPT12}. Again we observe that for larger $\lambda$-values (i.e.\ for $\lambda > 2$), that $k_{\rm c}$ depends only weakly on $\eta$ and the respective FMT-results are very close to the data originating from the LDA. For the smallest $\lambda$-value investigated, the results from the FMT deviate strongly from the LDA-results. Furthermore, we observe a pronounced $\eta$-dependence of $k_{\rm c}$. The origin of this $\eta$-dependency lies in the interplay between the purely entropic HC and the energetic soft shoulder contributions to the excess free energy. 

\begin{figure}[t!]
\centering
\includegraphics[width=0.5\textwidth]{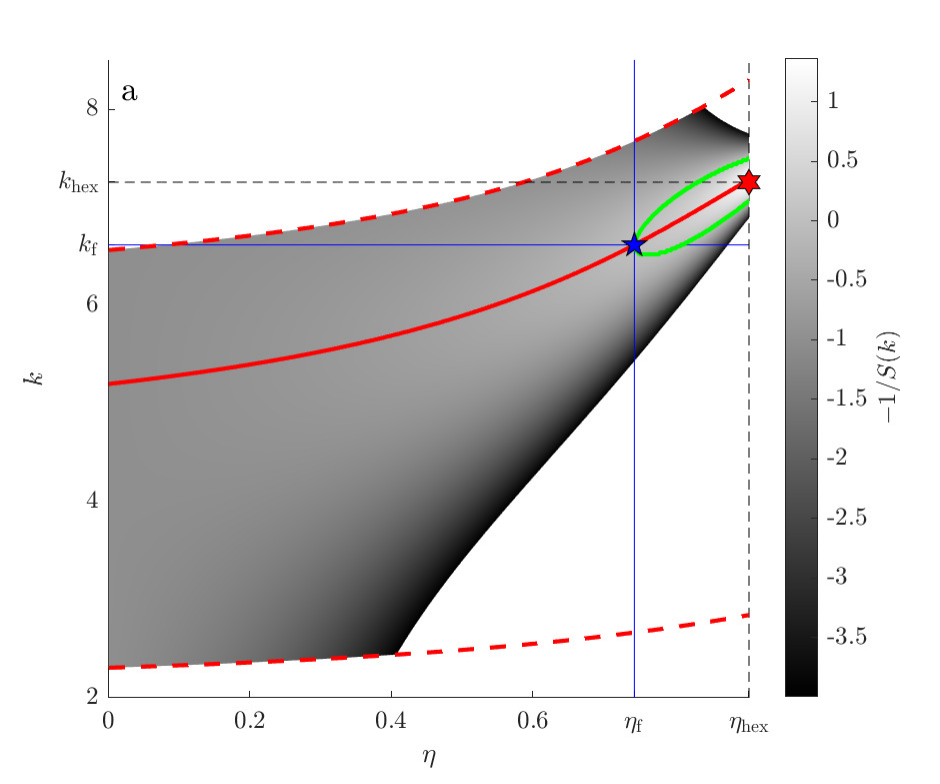}\includegraphics[width=0.5\textwidth]{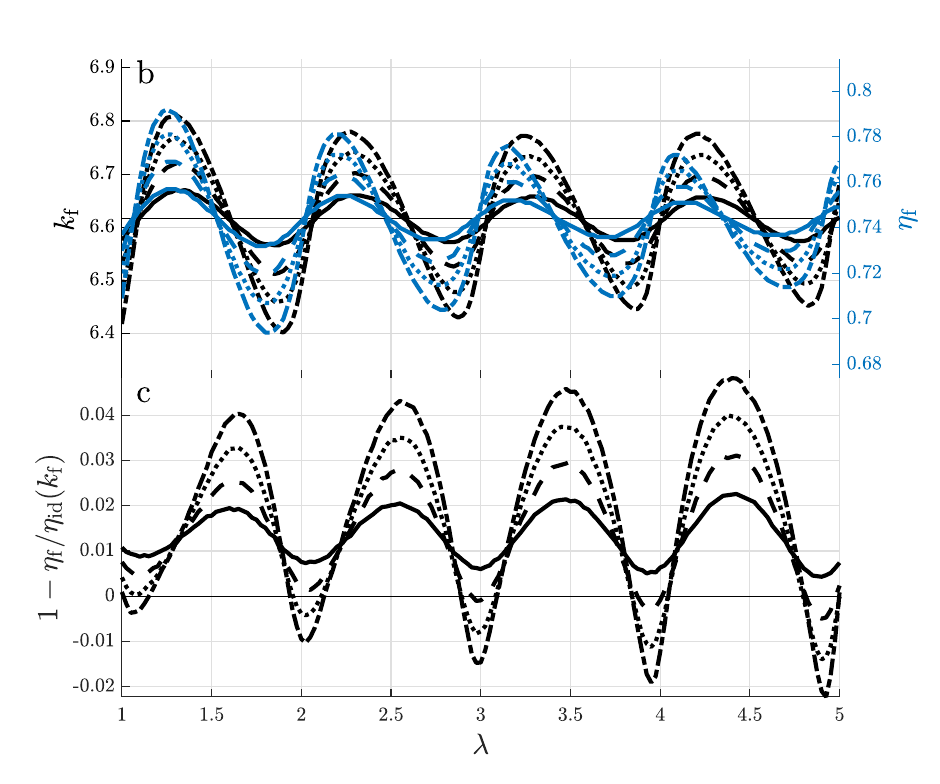}
\caption{{Panel a) shows a} heatmap plot of the negative reciprocal of the structure factor $-1/S(k)$ for the pure HC fluid as a function of $\eta$ and $k$, as obtained from FMT.
We employ a cutoff in the range displayed, $-1/S(k) \geq -4$, in order to {emphasise} those wavectors that predominantly govern the density modes in the pure HC fluid and determine the structure of the HC crystal. The positions of the first peak of $-1/S(k)$  (corresponding to the first peak in $S(k)$) is indicated by the red solid line. The red dashed lines mark the inflection points of the first peak of $S(k)$, i.e. where $\frac{d^2 S(k)}{d k^2}=0$. The distance between these dashed lines represents a measure for the width of the peak of $S(k)$. The blue star indicates the point in the $(\eta, k)$-plane where the HC liquid becomes marginally stable against $k_{\rm f}$, i.e.\ where $-1/S(k_{\rm f}) \to 0$ (see text). The red star marks the closest hexagonal packing at $\eta = 0.906$ and $k_{\rm hex} = 4\pi/\sqrt{3}$.
The green lines indicate the locus of the points where $-1/S(k)=0$.
{Panel b) displays a} plot of the freezing $k_{\rm f}$ as a function of $\lambda$ to show the effect of the shoulder interaction (in terms of the shoulder width $\lambda$) (black and left vertical axis) together with a plot of the freezing packing fraction $\eta_{\rm f}(\lambda)$ ({blue} and right vertical axis). These are for the scaled shoulder heights $\epsilon^* = \epsilon \lambda$ [see Eq.~\eqref{S_LDA}] of: $\epsilon^*=0.2$ (solid), $\epsilon^*=0.4$ (dashed), $\epsilon^*=0.6$ (dotted), $\epsilon^*=0.8$ (dash-dotted). The black horizontal line indicates the pure HC case. 
If we assume that at the freezing transition the particles form a perfect hexagonal lattice (governed by $k_{\rm f}$), one can derive an ideal packing fraction of such a lattice, given by $\eta_{\rm id}(k_{\rm f})=\sqrt{3}k_{\rm f}^2/32\pi$. The relative error between the predicted $\eta_{\rm f}$ and this `ideal' value $\eta_{\rm id}(k_{\rm f})$ is plotted in {panel c)} for the different $\epsilon^*$ values specified above.
}
\label{Fig4}
\end{figure}

To shed further light on how repulsive shoulder affects the HC freezing (with corresponding associated wavenumber $k_{\rm f}$), it is instructive to take a closer look on the DFT-predictions for the structure of the HCSS system in comparison to the pure HC liquid. In {Fig.~\ref{Fig4}a} we present a heatmap plot of $-1/S_{\rm FMT}(k)$ for the pure HC system (calculated with FMT) as a function of both $\eta$ and $k$. Cuts through this are also displayed as line plots in {Fig.~\ref{Fig3}a}. We have chosen a cutoff of $- 1/S(k) \ge -4$, to {emphasise} the dominant wavevectors, which at high $\eta$ determine the structure (lattice spacing) of the pure HC crystal, i.e.\ which are ``entropically favoured''.
In {Fig.~\ref{Fig4}a}, the red solid line denotes the position of the first maximum of $S(k)$ while the dashed red lines mark the inflection points of this function, defining thereby the width of the peak in $S(k)$ as the distance between these. The blue star marks the $\eta$-value for freezing ($\eta_{\rm f} = 0.744$ -- blue vertical line) of the system into a hexagonal crystal which is governed by the  $k_{\rm f} = 6.616$ (blue horizontal line).
The HC liquid is marginally stable for $k_{\rm f} \approx 6.616$ at $\eta_{\rm f}$.
As the density is increased, the distances in the hexagonal crystal shrink (with a corresponding increase of $k_{\rm f}$) until particles reach the closest hexagonal packing, occurring at $\eta \approx 0.906$ with $k_{\rm f}=k_{\rm hex}=4\pi/\sqrt{3}$ (see the red star in Fig.~\ref{Fig4}).
{We should emphasise that for the states bounded by the green curve in Fig.~\ref{Fig4}a, the structure factor of the fluid attains nonphysical negative values. Of course, this is a sign that the system is not a fluid at these state points and is in fact in a solid state.}
Note in {Fig.~\ref{Fig4}a} that even at intermediate values of the packing fraction (i.e., for $\eta \approx 0.5$) the range of favoured wavectors narrows down, approaching a fairly narrow range in the vicinity of the maximum (solid red line).

Having considered the pure HC system, we now move on to discuss the $\epsilon>0$ case.
Additional wavevectors are introduced by the shoulder interactions.
Those that lie in the peak region of the HC fluid are reinforced (amplified), especially at higher densities, while the smaller wavevectors are increasingly suppressed as the density is increased. For example, in the case where $\lambda=1.3$ discussed above, $k_{\rm c}\approx j_{2,1}/1.3 \approx 3.95$ so that it is in the peak region of the HC fluid up to $\eta\approx0.6$. This is why this is also roughly the position of the maximum in $k_{\rm c}(\eta)$, displayed in {Fig.~\ref{Fig3}c}.

In {Fig.~\ref{Fig4}b} we illustrate the impact of the square-shoulder interaction on the freezing transition.
This plot shows how $k_{\rm f}$ and $\eta_{\rm f}$ change as $\lambda$ is varied.
We compare the behaviours for different values of the scaled shoulder height $\epsilon^*=\lambda\epsilon$, showing results for $\epsilon^*\in\{ 0.2,0.4,0.6,0.8\}$.
Scaling in this way leads to comparable amplitudes in the shoulder contributions to $S(k)$ -- see the combination of coefficients in the second term of Eq.~\eqref{S_LDA} and also Fig.~\ref{Fig3}.
For large $\eta$ (i.e., $\eta>0.8$) only one (freezing) peak is observable in the HCSS $S(k)$, as long as $\epsilon$ remains reasonably small.
The freezing wavevector $k_{\rm f}$ displayed in {Fig.~\ref{Fig4}b} is determined by increasing $\eta$ until the global maximum in $-1/S(k)$ at this value reaches zero, i.e.\ satisfying $-1/S_{\rm FMT}(k_{\rm f})=0$. Of course, this also corresponds to the peak of the largest maximum of $S(k)$ diverging{, indicating the uniform liquid is unstable and so in reality freezing has already occurred at a lower density.
Nonetheless, it is instructive to note the value of $k_{\rm f}$ at which $S(k)$ diverges and to examine the behaviour with varying $\lambda$}.
We observe that both the predicted freezing $k_{\rm f}$ as well as the freezing packing fraction $\eta_{\rm f}$ are functions that oscillate around the HC value (shown as a black, horizontal line) as $\lambda$ increases.
Of all the maxima in $-1/S(k)$, the one that determines $k_{\rm f}$ depends on $\lambda$.
The role of which $k_i$ (i.e.\ which maxima of $S(k)$) becomes the freezing wavenumber $k_{\rm f}$ is passed from one to the next roughly at the $\lambda$ values corresponding to the position of the maxima in the oscillations of $k_{\rm f}(\lambda)$.
For example, at $\lambda=3.7$ it is the wavevector of the fourth maximum, i.e.\ $k_{\rm f}=k_4$.

Our MC simulations shows that, to a fairly good approximation, $k_{\rm f}$ governs the hexagonal lattice of hard cores that emerges when the liquid temperature is decreased at high densities.
However, there is a difference between the `ideal' packing fraction of the hexagonal lattice with wavevector $k_{\rm f}(\lambda)$ {(i.e.\ assuming that every vertex in this hexagonal lattice is occupied by a particle)} and the packing fraction $\eta_{\rm f}(\lambda)$ where $-1/(S(k_{\rm f})\to 0$. Moreover, this difference has consequence.
The ideal hexagonal lattice with this wavevector has packing fraction $\eta_{\rm id}$, determined as the ratio of the area occupied by a single particle and the area of the unit cell, giving $\eta_{\rm id}(k_{\rm f})=\sqrt{3}k_{\rm f}^2/(32\pi)$.
We display in {Fig.~\ref{Fig4}c} the relative difference between $\eta_{\rm id}(k_{\rm f})$ and $\eta_{\rm f}$.
We can interpret positive deviations of this quantity as an indicator of the number of vacancy defects in the crystal{, which Fig.~\ref{Fig4}c indicates can be rather large.
However, this comparison still does not tell us where the freezing transition occurs. This analysis indicates that freezing must occur at a density value lower than that where $-1/(S(k_{\rm f})\to 0$, but the precise density value comes from calculating the grand potential $\Omega$ of the liquid and crystalline phases and identifying phase coexistence in the usual way \cite{hansen13}.}

\subsection{The structure of the liquid state}
\label{subsec:liquid}

In the previous sections we showed that knowledge of the significant wavectors governing the structure of the HCSS liquid, $k_i$, $i = 1, \dots, 4$, provide valuable information about which structures can potentially emerge when the liquid becomes unstable and solid phases form. Further, we showed that the shoulder interactions lead to the introduction of additional wavevectors, beyond those of the pure HC system. The relevance of these is determined by their interplay with the core interactions, i.e.\ whether the nonlinear couplings in the free energy lead to a reinforcement of these modes or not. Density modes that are likely to be favoured can be identified in plots of $-1/S(k)$, such as that in Fig.~\ref{Fig3}. In this section we now bench-mark the FMT-RPA's ability to accurately describe the peaks of $S(k)$, on a quantitative level.

\begin{figure}[t!]
\centering
\includegraphics[width=1.0\textwidth]{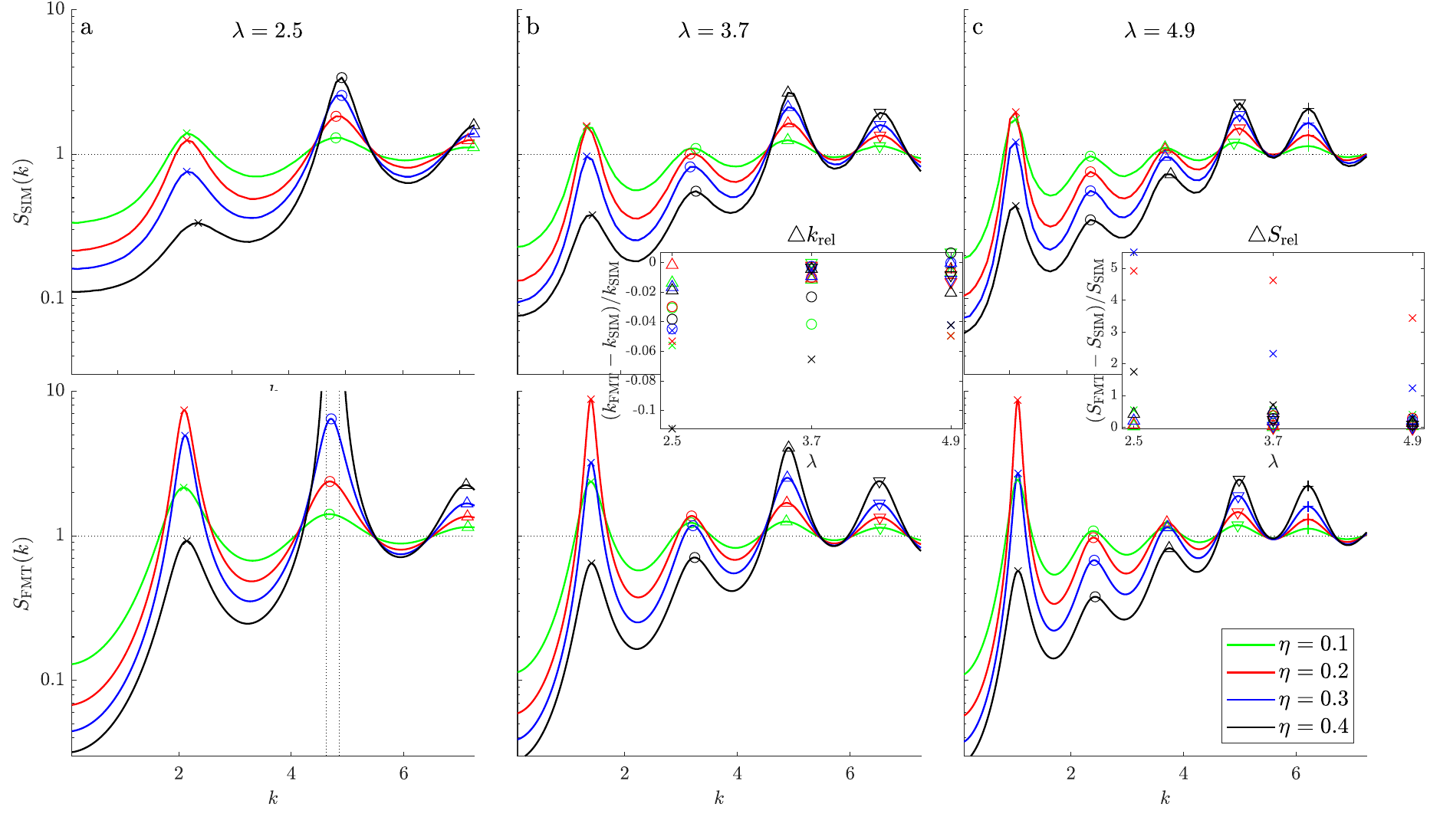}
\caption{Static structure factor $S(k)$ as a function of wavenumber $k$, as obtained in GCMC simulations (top panels) and from the FMT-RPA-DFT (bottom row), for different values of shoulder width, $\lambda$, and packing fractions, $\eta$ {(Panel a): $\lambda=2.5$, b): $\lambda=3.7$, c): $\lambda=4.9$)}. Note the semi-logarithmic scales. The temperatures $T$ were chosen to be slightly above the maximum temperature of the corresponding $\lambda$-line, $T_{\rm max}(\lambda)$: $T = 1.05\times T_{\rm max}(\lambda)$ -- see the points marked in {Fig.~\ref{Fig3}b}. Hence, these states are all above the $\lambda$-line, but lie very close for $\eta=0.2$ and $\eta=0.3$. For better visibility, the maxima $S_i$ of $S(k)$ are marked by symbols (crosses, circles, and triangles). The two insets show the relative errors between the simulation and DFT results for the peak positions, $k_i$, and peak heights, $S_i$, as functions of $\lambda$. In these insets the same symbols have been used for the different $S_i$ as in the plots of $S(k)$. These differences are defined as $\Delta k_{\rm rel} = (k_{\rm FMT}-k_{\rm SIM})/k_{\rm SIM}$ and $\Delta S_{\rm rel}=(S_{\rm FMT}-S_{\rm SIM})/S_{\rm SIM}$, respectively. A diverging second peak erroneously predicted by the FMT-RPA-DFT for $\lambda$ = 2.5 and $\eta = 0.4$ (see dotted vertical lines) has been exempted from these considerations.}
\label{Fig5}
\end{figure}

Thus, in Fig.~\ref{Fig5} we compare $S(k)$ obtained from GCMC simulations [via $g(r)$ and the inverse of Eq.~\eqref{ST2}] with the results obtained analytically via the FMT-RPA-DFT [Eqs.~\eqref{DFT14} and \eqref{c_FMT_hc}], for three different values of $\lambda$ (i.e., $\lambda$ = 2.5, 3.7, and 4.9) and four different values of $\eta$ (i.e., $\eta$ = 0.1, 0.2, 0.3, and 0.4). Note the semi-logarithmic axes in Fig.~\ref{Fig5}. We {have} chosen liquid states that are located just above the predicted $\lambda$-line (see Fig.~\ref{Fig3}), i.e., at a temperature $T = 1.05\times T_{\rm max}(\lambda)$, where $T_{\rm max}(\lambda)$ is the temperature for which the $\lambda$-line attains its maximum. The state-points at which these are calculated are marked in Fig.~\ref{Fig3}{b}. 

The first thing that we can observe in Fig.~\ref{Fig5} is that the number and positions of the individual peaks of $S(k)$ (i.e., the $k_i$), is in all cases predicted quite accurately by the FMT-RPA-DFT. A more quantitative measure is displayed the central inset, where we display the relative differences between the two sets of $k_i$, i.e., we plot $\Delta k_{\rm rel}=(k_{\rm FMT}-k_{\rm SIM})/k_{\rm SIM}$, for the three different values of $\lambda$. We find that the values of the $k_i$ are in general slightly underestimated by the DFT, with the largest error being for small $\lambda$- and higher $\eta$-values, notably for $\eta=0.4$ at $\lambda=2.5$.

\begin{figure}[t!]
\centering
\includegraphics[width=1.0\textwidth]{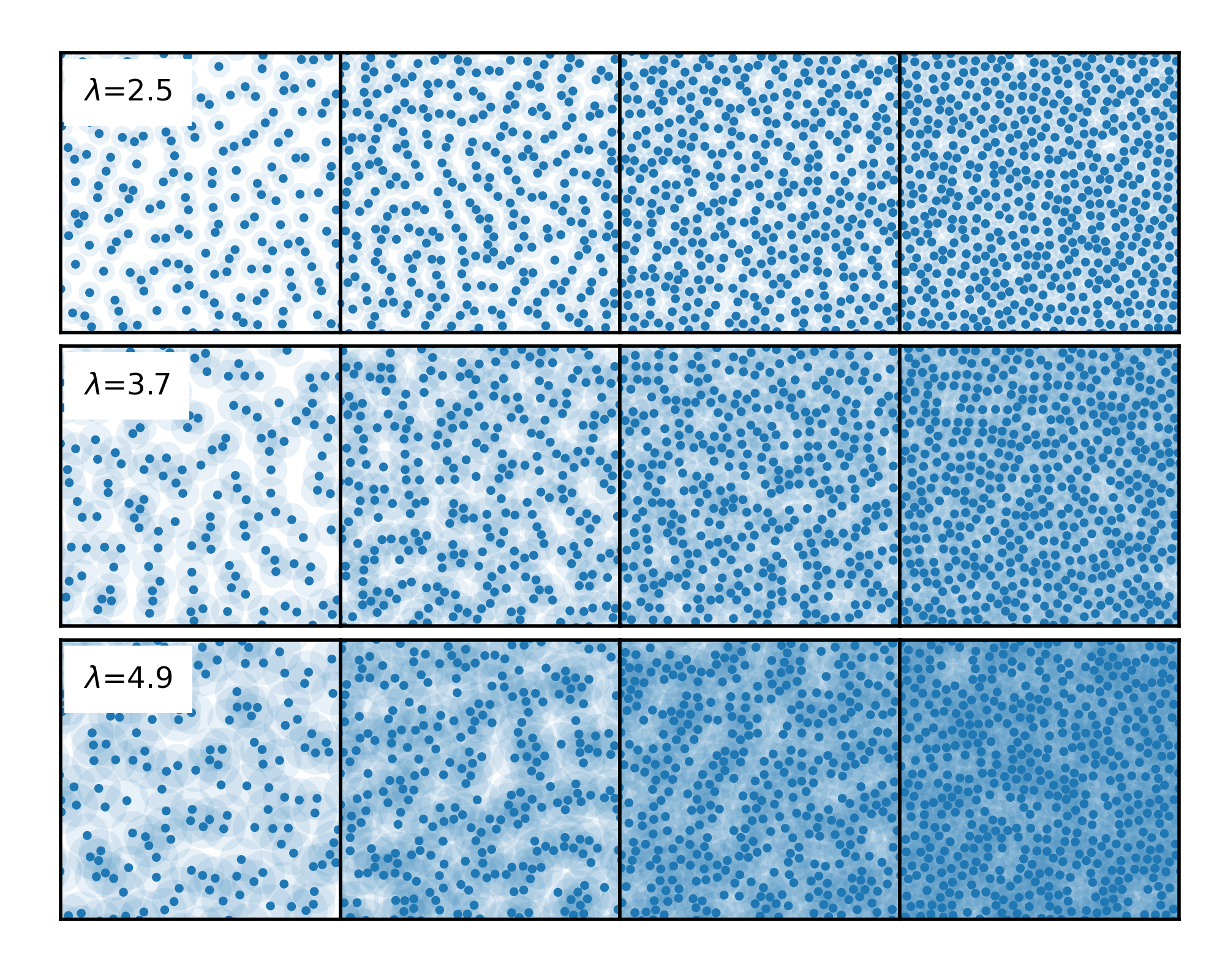}
\caption{Typical snapshots from GCMC simulations of all the states whose structure factors $S(k)$ are shown in Fig.~\ref{Fig5}. All systems are in the liquid phase. Snapshots are shown for $\lambda$ = 2.5 (top row), $\lambda$ = 3.7 (central row), and $\lambda$ = 4.9 (bottom row). The packing fractions are $\eta$ = 0.1, 0.2, 0.3, and 0.4 (from left to right).}
\label{Fig6}
\end{figure}

We also see in Fig.~\ref{Fig5} that the heights of the peaks $S_i$ are throughout overestimated by the DFT, in some cases only by a little and in other cases by a considerable amount. In the right hand inset of Fig.~\ref{Fig5}, we display the relative differences between the two sets of peak heights $S_i$, i.e., we plot $\Delta S_{\rm rel}=(S_{\rm FMT}-S_{\rm SIM})/S_{\rm SIM}$, for the three different values of $\lambda$. 
We see that the FMT-RPA-DFT greatly overestimates $S_1$ for states in the vicinity of the $\lambda$-line, but is relatively accurate for the remaining $S_i$.
This is particularly the case for $\eta=0.2$ and $\eta=0.3$, in the vicinity of the $\lambda$-line. In contrast to the DFT, the simulations show a tendency to suppress the first peak with increasing $\eta$. For $\lambda = 2.5$ and $\eta=0.4$ the DFT predicts a very large (in fact diverging) value of $S_2$, indicating that the liquid is linearly unstable against $k_2$. Note that this $k_2$ neither corresponds to $k_{\rm c}$ nor to $k_{\rm f}$. What is also interesting about this peak is that for intermediate densities and for $\lambda \gtrsim 2.4$, it points to the possibility of the formation of structures with characteristic wavelength that is neither that associated with the core- nor with the shoulder-length scale. This observation tallies with those in Ref.~\cite{Fornleitner_2010}, where low-density crystals were identified that are compatible with these observations.

Despite the DFT prediction of structured phases inside the $\lambda$-line (e.g.\ for $\lambda=2.5$ and $\eta=0.4$), all the states investigated within our simulations at the temperatures considered in Fig.~\ref{Fig5} are in fact in the liquid phase. This can be seen from the corresponding simulation snapshots displayed in Fig.~\ref{Fig6}.
Nonetheless, these snapshots do show significant structuring in the liquid.
The characteristic wavelengths of these patterns are predicted accurately by the DFT, because the DFT predictions for the peak positions in $S(k)$ are correct.
It is at lower temperatures where the system finally freezes into a variety of different solid phases.
{In other words, the DFT predicts the transition to the structured phases at temperatures higher than they are observed in the MC simulations. However, we should {emphasise} that even within the DFT description, one must expect that coexistence between the liquid and the ordered phases to occur before the structure factor diverges. Thus, the DFT is self-consistent, with the structure factor not actually diverging at the transition -- except perhaps at the top of the $\lambda$-line itself \cite{archer2007model, Archer2008, Archer2008a}.}

In the following section we discuss the simplest and most commonly observed of these {structured} phases, namely the cluster, stripe and hole phases.
A variety of other complex ordered phases can also arise, but a detailed investigation of these will be presented elsewhere.
For higher temperature state points further away from the $\lambda$-line, the agreement between the DFT and simulations for the values of $S_i$ is considerably better.

\subsection{Structured phases}
\label{subsec:solids}

\begin{figure}[t!]
\centering
\includegraphics[width=1.0\textwidth]{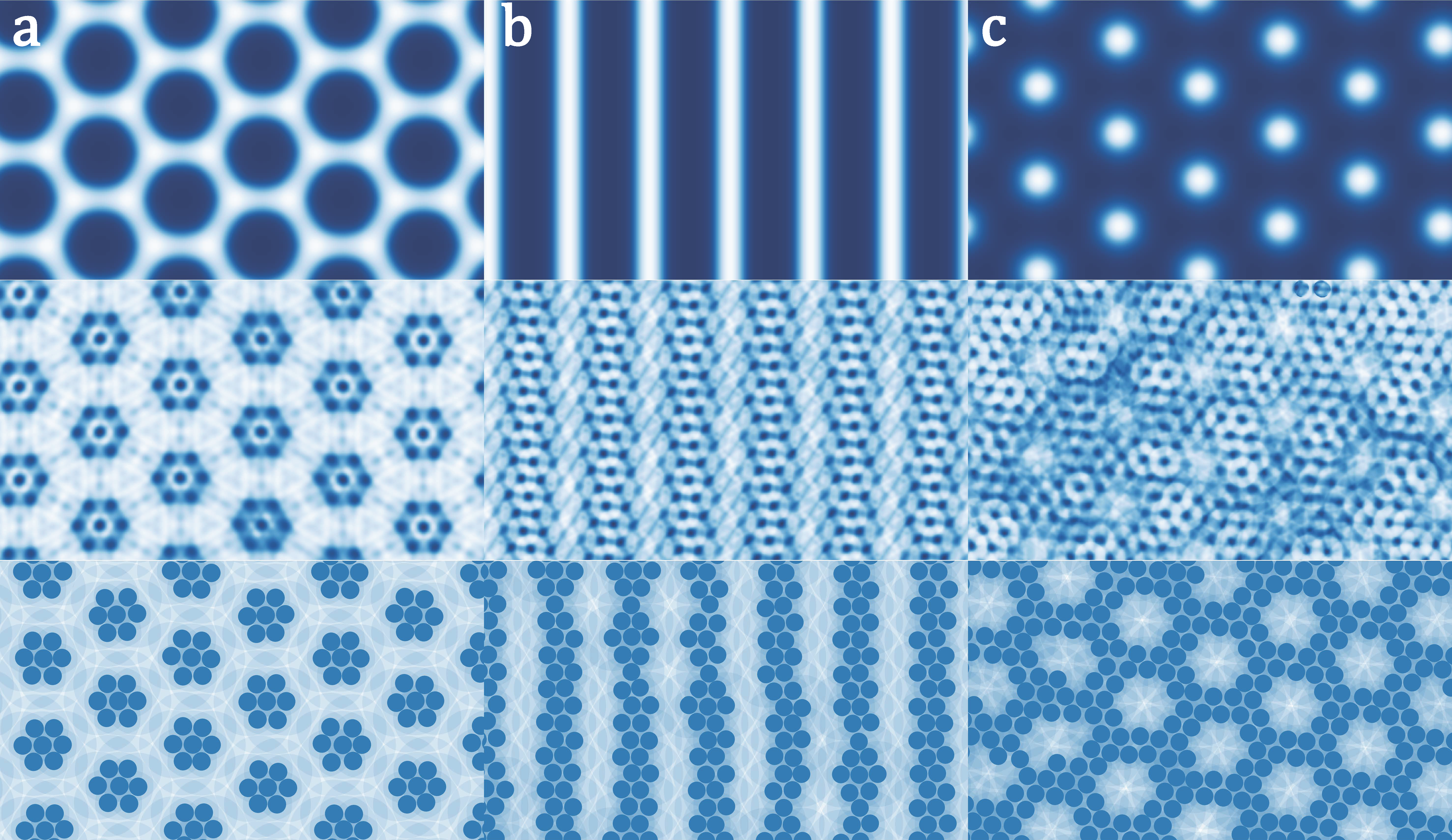}
\includegraphics[width=1.0\textwidth]{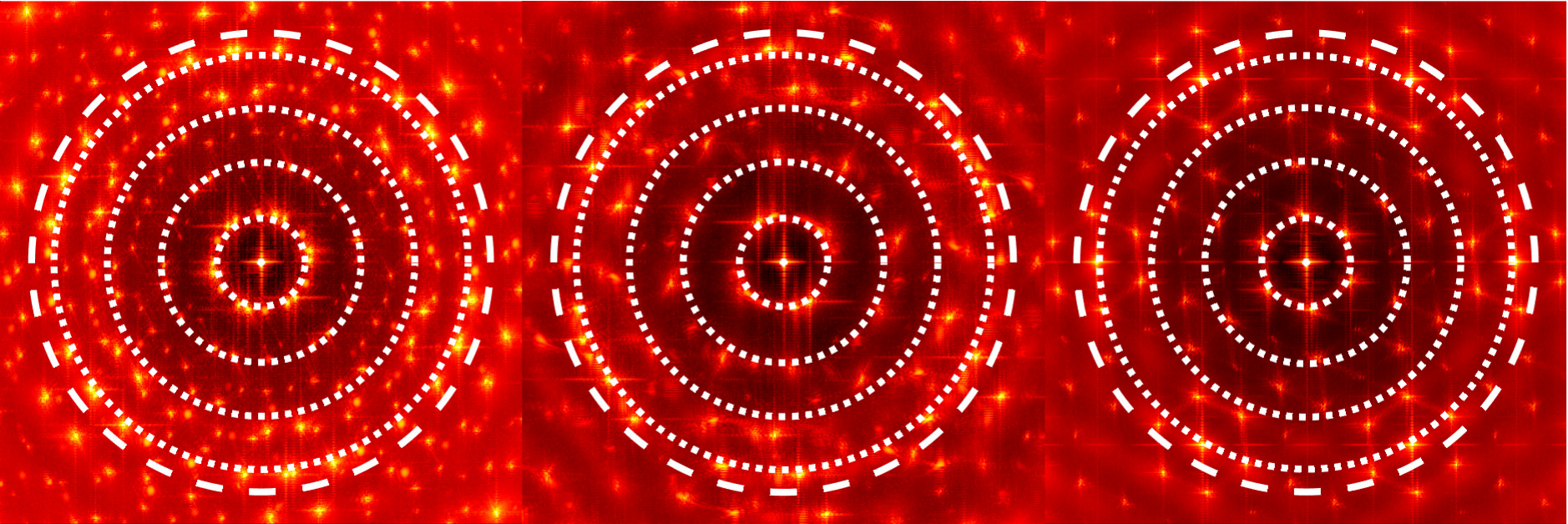}
\caption{Comparison of equilibrium density profiles $\rho({\bf r})$ for different HCSS-systems (specified below) obtained via the LDA-RPA-DFT (first row), the FMT-RPA-DFT (second row), and equal-sized snapshots taken from GCMC simulations of much larger sized systems (third row). The bottom panels show the corresponding structure factors $S({\bf k})$  as functions of ${\bf k} = (k_x, k_y)$ extracted from GCMC simulations. The white circles correspond to the peak positions of $S(k)$ observed via the FMT-RPA-DFT for a bulk liquid at the same $T$, and $\mu$ values. The columns show {a):} a {cluster} phase for $\mu(\eta=0.25)$ [see Eq.~\eqref{LS_mu}] and $T=0.25$; {b):} a {striped} phase for $\mu(\eta=0.35)$ and $T=0.25$; and {c):} a holed phase for $\mu(\eta=0.45)$ and $T=0.38$.}
\label{Fig7}
\end{figure}

So far, we have largely used the DFT to determine the liquid state structure and to show that it predicts fairly well the key wavenumbers $k_i$.
In this section, we now discuss the DFT predictions for properties of structured phases arising in those parts of the phase diagram where both the DFT and the simulations predict structured phases.
We focus in particular on comparing wavelengths of density modulations and assessing how the DFT fares in predicting these, in comparison with the simulations.
Comparing the equilibrium density profile $\rho({\bf r})$ obtained from full {minimisation} of both the LDA- and the FMT-RPA functionals with data obtained in GCMC simulations, {we find that at higher temperatures, both DFTs overestimate the size of the regions in the phase diagram where the periodic phases exist, but overall are in good qualitative agreement with the MC results.
The LDA-RPA-DFT predicts a phase diagram that is qualitatively very similar to that of the SALAR system displayed in Ref.~\cite{Archer2008a}.
However, at lower temperatures the LDA is unable to predict the freezing of the particles that occurs within the clusters and stripes of the modulated phases, nor can it describe the variety of other crystalline phases that can arise at lower temperatures and/or higher densities.
As illustrated in Fig.~\ref{Fig7}, the FMT-RPA-DFT does fare better for the cluster, stripe and hole phases.
Additionally, at lower temperatures and/or higher densities, in addition to these three periodic phases, a variety of other crystalline phases also occur, depending on the value of $\lambda$.
But, as mentioned, detailed investigation of these other phases will be presented elsewhere.
Fig.~\ref{Fig7} shows some examples of density profiles obtained at lower densities, where we find good agreement between the FMT-RPA-DFT and the GCMC simulations, although}
there are some errors made by the DFT that we show are introduced by the underestimation of the $k_i$ discussed in the previous section.

As discussed in the previous section, due to the overestimation of the height of the first peak $S_1$, the RPA-based DFT functionals tend to predict the formation of patterned structures a temperatures higher than they actually occur, i.e., within the $\lambda$-line, which is at too high temperatures. For instance, the clustering transitions discussed in Sec.~\ref{subsec:liquidcluster}, is found using GCMC simulations at $\eta \approx 0.2$ to occur at $T_{\rm c} \approx 0.36$ (see Fig.~\ref{Fig1}), whereas the $\lambda$-line for the same $\eta$-value is found using the LDA at $T_{\rm c}=0.6181$ and from the FMT it is at $T_{\rm c}=0.6979$. Similarly, at a higher density, using GEMC simulations we find that the transition from the liquid to the cluster phase occurs at  $\eta \approx 0.2827$ and $T_{\rm c} \approx 0.3663$, whereas for this density the LDA predicts $T_{\rm c}=0.5892$ and the FMT at $T_{\rm c}=0.6829$. 
Thus, to compare predictions from the DFTs with simulation results for the structured phases, we must choose state points at relatively low temperatures, where both the LDA- and FMT-RPA-DFT reliably predict the emerging phases. 

As was discussed in Sec.~\ref{subsec:liquid}, at higher densities and for higher values of $k$, the FMT-RPA functional tends to be more accurate in its predictions for the $S_i$.
Hence, in such situations we can expect this functional to be more reliable for predicting the phase transitions associated with the corresponding $k_i$.
Despite the inaccuracy of the DFTs for the values of the peak heights $S_1$, we find the DFT still accurately predict the equilibrium structures. This can be seen in Fig.~\ref{Fig7}, where we compare density profiles for the striped, clustered, and holed phases (other phases can also be observed in some cases; these will be presented elsewhere), together with corresponding simulation snapshots.
In the top row, we display equilibrium density profiles $\rho({\bf r})$ obtained from the LDA-RPA-DFT. Note that for better visibility of the profiles in the {low-density} regions, we plot the logarithm of the density profile. Unsurprisingly, given that these results are from a functional that treats the hard-disk contributions via the simplest LDA-approximation, this DFT is only able to resolve the large-scale structuring, {characterised} by the length-scale $a_c$, induced by the shoulder interactions, which are treated within the RPA approximation.

The second row of Fig.~\ref{Fig7} displays $\rho({\bf r})$ obtained from the FMT-RPA-DFT. These are obtained by using the LDA-RPA density profile as the initial guess in the numerical procedure for {minimising} the FMT-RPA functional.
Note that results calculated in this manner were found to be consistent with those obtained by {starting} from a random initial guess for the density profile. However this `educated guess', based on the LDA-RPA profile, was found out to be computationally more efficient, since both density profiles share similar large-scale features, as they use the same RPA treatment of the shoulder.
We see in Fig.~\ref{Fig7} that the density-profile $\rho({\bf r})$ obtained from the FMT-RPA has the same stripe/cluster/hole interspacing as the LDA-RPA, but additionally resolves the hexagonal ordering formed by the particles being frozen within the sub-structures.
{At first glance, the FMT-RPA results in the second row appear somewhat `messy'. However, understanding of what these show can be obtained by comparing the FMT-RPA results with equal-sized snapshots extracted from GCMC simulations, which are displayed in the third row.
Recall that DFT is a statistical mechanical theory, calculating the ensemble average density profile \cite{Evans1979, hansen13}.
Thus, comparing the second and third rows, one can see that the second row FMT-RPA profiles do indeed resemble ensemble averages of configurations like those displayed in the third row.
Additionally, from comparing the DFT profiles with the respective GCMC snapshots we also can see that while the local hexagonal order, governed by $k_4$,} is very well predicted by the FMT-RPA, both the LDA-RPA and FMT-RPA overestimate the larger length-scale associated with $k_1=k_{\rm c}$. This can most easily be seen by comparing the spacing between stripes in {the three upper panels of Fig.~\ref{Fig7}b}.

\begin{table}[t!]
    \centering
    \begin{tabular}{|c|c|c|c|c|} \hline
         phase & $\eta_{\rm b}$ & LDA-RPA & FMT-RPA & GCMC \\ \hline
        clusters & 0.25 & 0.2499 & 0.2517 & 0.2883 \\ \hline
         stripes & 0.35 & 0.3466 & 0.3636 & 0.3905 \\ \hline 
         holes  & 0.45 & 0.4441 & 0.4640 & 0.4875  \\ \hline
    \end{tabular}
    \caption{Average packing fractions of the three investigated solid phases displayed in Fig.~\ref{Fig7}, obtained via DFT from GCMC simulations. For both the DFT and MC calculations, a fixed chemical potential $\mu(\eta_{\rm b})$ determined via Eq.~\eqref{LS_mu} was imposed.}
    \label{tab:solid_densities}
\end{table}

The average packing fractions of the structures displayed in Fig.~\ref{Fig7}, obtained by the DFT and also from the GCMC simulations using the same chemical potential values $\mu(\eta_{\rm b})$ [see Eq.~\eqref{LS_mu}] are given in Tab.~\ref{tab:solid_densities}.
We see that the LDA has the tendency to slightly underestimate the densities of the structured phases and to {predict} their occurrence at even lower densities, as compared to a bulk liquid with the same value of $\mu$.
In contrast, the FMT predicts an increase of the density compared to the reference unstable liquid wih imposed packing {fraction} $\eta_{\rm b}$. This is consistent with, but slightly less than the values obtained from the GCMC simulations.
Note that for simulations carried out at constant $\mu$, the increase of the density relative to the density of a bulk liquid with equal $\mu$ can be used as an indicator for the formation of a structured phase. This can also be seen in the Appendix, where a plot showing comparison of the bond order parameters and of $\eta$ for GCMC simulations is displayed.   

In the bottom {panels} of Fig.~\ref{Fig7} we display the structure factors $S({\bf k})$ obtained from the GCMC simulations. Similar to in Fig.~\ref{Fig2}, we have added on top of these heatmap plots the relevant significant wavectors as predicted by the FMT-RPA-DFT, obtained for a reference bulk liquid with the same $T$ and $\mu$. We find that -- even though the structured phases are found well below their freezing temperatures -- the wavenumbers extracted from the liquid theory still agree very well with the Bragg-peaks in $S({\bf k})$, especially for $k_1$ and $k_4$ (similar to in Fig.~\ref{Fig2}), but slightly underestimating the other wavectors, in accordance with observations made in Sec.~\ref{subsec:liquid}.
Note that the full simulation box from which these $S({\bf k})$ are obtained contains several distinctly oriented domains, leading to some overlay in the resulting Bragg-peaks. Nonetheless, we observe that the major discrete peaks in the GCMC $S({\bf k})$ are located very closely to the predicted $k_{\rm c}=k_1$ and $k_{\rm f}=k_4$ in $S_{FMT}(k)$.

Regarding both DFTs, the slight underestimation of the smaller wavenumber $k_1=k_{\rm c}$, leads to an overestimation of the interspacing between stripes/clusters/holes.
While the error made for $k_4$ has seemingly minimal effects on the predicted equilibrium density profile, the error in $k_1$ results in an overestimation of $a_{\rm c}$, hence the discrepancy in the spacings of the different structures in Fig.~\ref{Fig7} when comparing DFT and simulation. Although the error in $k_1$ is relatively small, due to the reciprocal relation between real- and Fourier-space, any small error $\delta k$, made in the determination of a wavevector $k$, leads to an error $\delta a \approx \frac{2\pi}{k+\delta k}-\frac{2\pi}{k}$ in the related length-scale of the real-space equilibrium density profile, that can be much more noticeable.
In other words, errors made for smaller $k$ lead to more severe consequences than the ones for bigger wavevectors.
Nonetheless, we conclude that that we can use the liquid state DFT based theory to successfully predict the wavevectors shaping the various solid states.

\section{Concluding remarks}
\label{sec:conc}

In this contribution we have investigated in detail the pattern formation of a two-dimensional {hard-core, square-shoulder} (HCSS) system at the point where it becomes unstable with respect to the emerging ordered phases.     The interaction potential acting between the particles is {characterised} by a circular, impenetrable core with an adjacent, repulsive square shoulder interaction; the latter is {characterised} by the step height (which can be treated as the inverse of the system temperature) and the range of the interaction, $\lambda$.  Investigations were carried out via intensive Monte Carlo simulations (in the grand-canonical -- GCMC -- and in the Gibbs ensemble -- GEMC) and via classical density functional theory; in the latter case two different approximations were implemented for the hard core part of the potential (based on the local density approximation -- LDA -- and on fundamental measure theory -- FMT), while the shoulder part of the potential was included in a mean-field type random phase approximation (RPA). 

Our simulation-based investigations have provided evidence that the density modes that dominate fluctuations in the liquid state (in terms of relevant wavevectors) of the HCSS system are the ones that ultimately shape the adjacent (ordered) cluster phase, once the liquid becomes unstable. We have elucidated how the interplay of the characteristic length scales of the {hard-core} and soft-shoulder interactions give rise to the dominance of specific density modes.
We then have checked the capacities of the two different DFTs for predicting the relevant wavevectors; we find that they agree with the MC simulations reasonably well.
Focusing on the liquid phase, we have compared the static structure factor as obtained directly from the simulations with data obtained via DFT: we observe that the positions of the peaks in the static structure factor agree very nicely, while larger deviations between simulation- and DFT-results are observed for the peak heights, in particular in the vicinity of the $\lambda$-line.

We have also investigated the structured phases and addressed the question: in what respect do the aforementioned differences between simulation and DFT results affect the density profiles predicted by DFT the ordered phases? We conclude from our investigations that errors in the small-$k$ regime have a larger impact than deviations occurring for larger wavevectors. However, we consider FMT to be sufficiently accurate to predict the significant wavevectors. Thus it seems a valid approach to apply DFT to the liquid phase, in order to reliably predict those wavevectors that {characterise} the emerging ordered phases. 
We conclude that the DFTs developed here can be used in a rather simple manner to navigate (with a relatively small computational cost) through the complex phase diagram of the HCSS system.
In return, this feature might also enable us to design interaction potentials that give rise to specific combinations of density modes in emerging, self-assembled ordered phases.

It should be noted that observations similar to the ones here have been made for liquid metals \cite{Hafner_1984}. Using effective interatomic pair potentials and reliable liquid state theories \cite{Kahl_1984} to calculate the static structure factor of the liquid phase, good {agreement} to the well-known crystalline structures of the metals was {observed}. 

\section*{Acknowledgments}

The computational results presented here were enabled via a generous share of CPU time, offered by the Vienna Scientific Cluster (VSC) under project number \#71263. Particular thanks is due to Ms.~Katrin Muck for her guidance related to the use of HPC. AJA gratefully acknowledges support from the EPSRC under grant EP/P015689/1.

\appendix
\section*{Appendix}
\label{appendix}

\subsection{{Numerical solution of the DFT -- Picard algorithm}}
\label{subsec:Picard}

The numerical {minimisation} of the DFT functionals used in this contribution via the iterative Picard algorithm \cite{Roth2010, hughes2014introduction} allows for the numerical determination of the equilibrium density profile $\rho({\bf r})$.
In the following we briefly outline the procedure.

{Minimisation} of the grand-potential functional~\eqref{DFT2},
\begin{equation}
\frac{\delta \Omega[\rho]}{\delta \rho ({\bf r})}=0,
\end{equation}
leads (in the absence of an external potential) to the Euler-Lagrange equation
\begin{equation}\label{PIC1}
\log{(\rho({\bf r}))}-c^{(1)}({\bf r})-\mu=0,
\end{equation}
with $c^{(1)}({\bf r})\equiv-\delta{\cal F[\rho]}_{\rm ex}/\delta\rho ({\bf r})$ being the one-body direct correlation function \cite{hansen13}. Evaluating the latter for a uniform liquid allows us to express {this reference} bulk density $\rho_{\rm b}$ in terms of the chemical potential $\mu$,
\begin{equation}
\rho_{\rm b}=e^{-\mu+c^{(1)}|_{\rho_{\rm b}}}.
\end{equation}
With this relation at hand we can rearrange Eq.~\eqref{PIC1} to obtain the following expression for the density profile:
\begin{equation}\label{PIC_guess}
\rho_{\rm P}({\bf r})=\rho_{\rm b} e^{c^{(1)}({\bf r})-c^{(1)}|_{\rho_{\rm b}}}.
\end{equation}
The index ``P'' indicates that this expression can be used in the iterative algorithm to obtain an update for the density profile at a given step, when the density profile of the preceding step is inserted into the right hand side of Eq.~\eqref{PIC_guess}. {It should be {emphasised} that $\rho_{\rm b}$ is only equal to the average density of the system $\bar{\rho}=\langle\rho({\bf r})\rangle$ for uniform (fluid) states.
More generally, in the case of the non-uniform phases, the average density $\bar{\rho}\neq\rho_{\rm b}$, although differences can be small. Thus, Eq.~\eqref{PIC_guess}} represents the basis for the iterative Picard algorithm as described e.g.\ in Refs.~\cite{Roth2010, Archer2015}.

To be more specific, we start from an initial guess for $\rho({\bf r})$ and then a new density profile is constructed by mixing the density profile of the preceding step via a mixing parameter $\alpha$ with the density profile $\rho_{\rm P}({\bf r})$, given Eq.~\eqref{PIC_guess}. For reasons of numerical stability one must introduce a small value for the mixing {parameter $\alpha$.}
Thus the density profile at step $(i+1)$ of the Picard iteration is given by
\begin{equation}\label{PIC_mix}
\rho^{(i+1)}({\bf r})=\alpha\rho_P({\bf r})+(1-\alpha)\rho^{(i)}({\bf r}).
\end{equation}
This algorithm is iterated until the squared difference between the density profiles of two consecutive steps integrated over the domain, i.e, if the dimensionless quantity $\sigma^2\int|\rho^{(i)}({\bf r})-\rho^{(i-1)}({\bf r})|^2 d{\bf r}$, drops below a predefined threshold value. The actual value depends on the system size, but typically amounts to $(10^{-10}-10^{-14})$. Simultaneously, the grand potential $\Omega[\rho]$ converges towards a constant value.

The mixing parameter $\alpha$ is chosen to be small enough to ensure convergence of the sequence of density profiles emerging from Eq.~\eqref{PIC_mix}. In practice this value is chosen by trial; in the case of the LDA-RPA functional a value of $\alpha=10^{-3}-10^{-2}$ was sufficient to obtain convergence, depending on initial density $\rho_{\rm b}$ as well as on the value of the shoulder height $\epsilon$. For the FMT-RPA functional we implemented an algorithm proposed by Roth in \cite{Roth2010} that evaluates the optimal choice of $\alpha$ in each step.

\subsection{Structure factor}
\label{A1}

To calculate the static structure factor, in addition to using the method described in Sec.~\ref{subsec:liquidcluster}, we also used the more standard definition \cite{hansen13}:
\begin{equation}\label{ST4}
S({\bf k})=1+\left\langle \frac{1}{N}\sum_{i=1}^N\sum_{j\neq i}^N e^{-i{\bf k}\cdot({\bf r_i}-{\bf r_j})}  \right\rangle.
\end{equation}
This relation requires as input the positions of all the particles $\{{\bf r}_i\}$, obtained in our MC simulations. The brackets denote an ensemble average in the related ensemble (here the grand-canonical or Gibbs ensemble). 

We also compared results for $S(k)$ obtained via Eq.~\eqref{ST4} with the numerical calculation of $S(k)$ via the inverse Fourier transform of Eq.~\eqref{ST2}, starting from the $g(r)$ obtained in GCMC simulations (comparison not displayed), since the latter can be prone to numerical errors. This is because the two-dimensional Fourier transform requires special care when {discretising} $k$, in view of the resolution in $r$-space. The results for $S(k)$ obtained via the two routes were found to be in good agreement, except for in the (irrelevant for present purposes) small-$k$ region, for $k<k_1$.

\subsection{Identification of the liquid-cluster transition temperature}

Here, we provide evidence of the approximate value of the transition temperature $T_{\rm c}$ between the liquid and the ordered cluster phase, for the HCSS system with $\lambda = 3.7$ and chemical potential $\mu(\eta = 0.2)$, calculated via Eq.~\eqref{LS_mu}. 

\begin{figure}[t!]
\centering
\includegraphics[width=0.5 \textwidth]{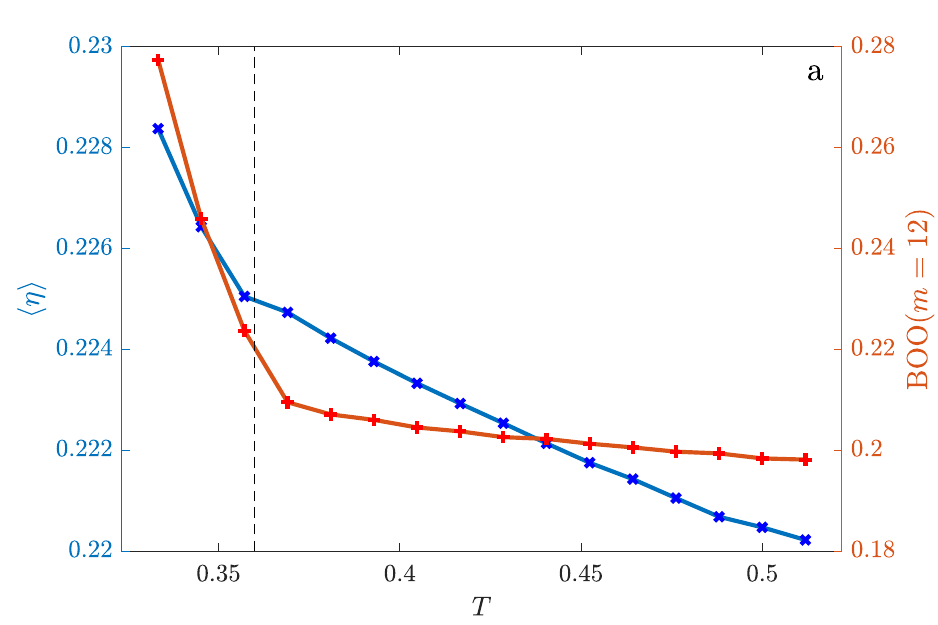}\includegraphics[width=0.5 \textwidth]{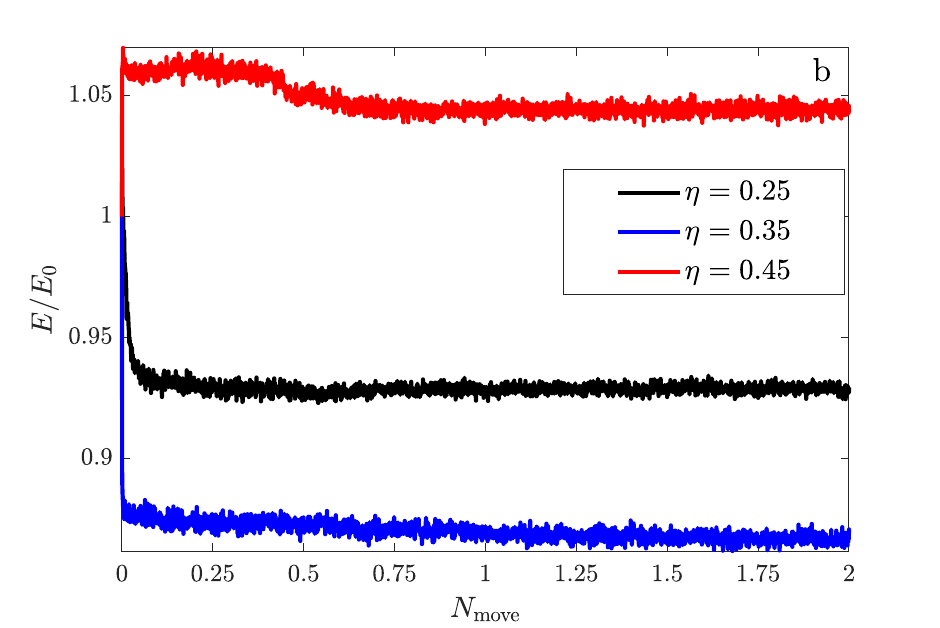}
\caption{{Panel a)}: GCMC results for ensemble averaged packing fraction $\langle \eta \rangle$ (blue symbols and line; left coordinate) and the BOO$(m = 12$) (with particles being in shoulder contact) as functions of temperature $T$; the system is {characterised} by $\mu(\eta=0.2)$ -- see Eq.~\eqref{LS_mu} -- and $\lambda = 3.7$. The temperature range considered covers the transition between the liquid and the ordered cluster phase. The dotted, black line marks our estimate for the transition temperature $T_{\rm c} \approx 0.36$. {Panel b)}: evolution of the energy of the system ({normalised} by the energy of the initial configuration) as a function of MC moves as obtained in GCMC simulation of the ordered states depicted in Fig.~\ref{Fig7}.}
\label{SupportGCMC}
\end{figure}

GCMC simulation have been performed, gathering data for the ensemble averaged packing fraction $\eta$ and the bond orientational order parameter BOO$(m = 12)$. The results are shown in Fig.~\ref{SupportGCMC}. For temperatures above $T \approx 0.36$ we observe a rather flat BOO$(m = 12)$ and only moderately increasing average packing fraction $\langle \eta \rangle$ with decreasing temperature. In contrast, both curves show a pronounced and characteristic increase as we pass the $T \approx 0.36$ threshold value from above. Thus we identify $T = T_{\rm c} \approx 0.36$ as a rather reliable estimate for the transition temperature between the two phases for this value of $\mu$. Note that the calculated packing fraction is slightly above the corresponding $\eta$-value of the bulk liquid calculated within SPT.  

In {Fig.~\ref{SupportGCMC}b} we display the evolution of the system energy, i.e.\ the number of shoulder overlaps, in the course of the GCMC simulation. The energy is divided by the energy of the initial hexagonal configuration of particles (see Sec.~\ref{subsec:simulations} for detail). After the rapid initial decrease (or increase in the case of the holed phase at $\eta_{\rm b}=0.45$), we observe equilibration of all states towards constant values.
It is hence safe to assume that our solid state simulations are equilibrated after about $10^{10}$ moves. For the ensemble averages, states from the end of the simulation are used.

\subsection{Liquid-cluster phase coexistence in the Gibbs-ensemble}

\begin{figure}[t!]
\centering
\includegraphics[width=0.5 \textwidth]{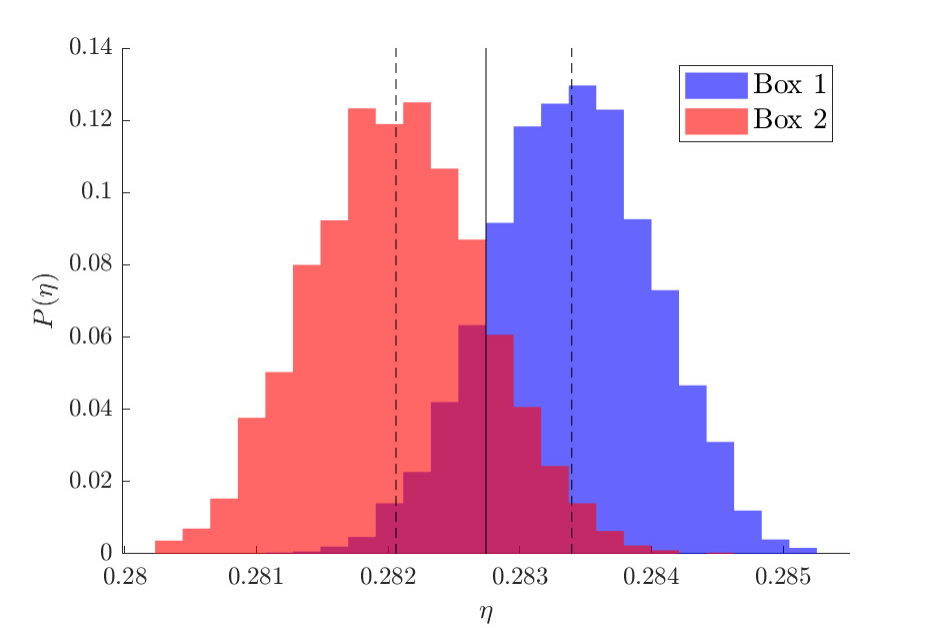}
\caption{Probability to find the two individual simulation boxes of the GEMC simulations (see Fig.~\ref{Fig2}) at a specific average packing fraction $\eta$.  The solid line denotes the (constant) packing fraction of the full Gibbs ensemble ($\eta=0.2827$) and the dashed line the average packing fraction in the individual boxes, respectively. Here, `Box 1' corresponds to the simulation box hosting the cluster phase ($\langle{\eta_{\rm cl}}\rangle=0.2834$), while `Box 2' hosts the liquid phase ($\langle{\eta_{\rm liq}}\rangle=0.2821$). }
\label{SupportGEMC}
\end{figure}

The extent of the liquid-cluster coexistence region can be obtained by calculating the average densities in the two simulation boxes used in the GEMC scheme.
Thus, we calculate the probability distributions $P(\eta)$ for each of the boxes to have the average packing fraction $\eta$.
Having these $P(\eta)$ form two distinct Gaussian distributions, can be seen as sign of proficient sampling of the Gibbs ensemble.
The GEMC simulation of the state at $\eta=2.827$ and $T= 0.3663$ involving $N=3000$ particles was run for $8\times10^9$ MC-moves. The statistics were obtained using 3000 states from the end of the simulation, from after phase separation in boxes {occurred}.
All sample states are separated by a single simulation block (see Sec.~\ref{subsec:simulations} for details). 
$P(\eta)$ and the corresponding densities of the liquid and clustered phase can be found in Fig.~\ref{SupportGEMC}.
Despite the small extent of the coexistence region, of width $\triangle \eta=\langle{\eta_{\rm cl}}\rangle-\langle{\eta_{\rm liq}}\rangle=0.0126$, the GEMC algorithm is capable of directly identifying phase coexistence.


%

\end{document}